\documentclass[draftcls, onecolumn]{IEEEtran}

\usepackage{amsmath}
\usepackage{amssymb}
\usepackage{color}
\usepackage{multirow}
\usepackage{array}
\usepackage{graphicx}
\usepackage{float}
\usepackage{cite}

\newtheorem{theorem}{Theorem}
\newtheorem{corollary}{Corollary}
\newtheorem{lemma}{Lemma}

\newtheorem{example}{Example}

\allowdisplaybreaks

\begin{document}
%
\title{Role of a Relay in Bursty  Multiple Access Channels}
%
%
%

\author{Sunghyun~Kim,~\IEEEmembership{Member,~IEEE,}
        Soheil~Mohajer,~\IEEEmembership{Member,~IEEE,}
        and~Changho~Suh,~\IEEEmembership{Member,~IEEE}
\thanks{S. Kim is with Electronics and Telecommunications Research Institute, South Korea. E-mail: {koishkim@etri.re.kr}.}
\thanks{S. Mohajer is with the Department of Electrical and Computer Engineering, University of Minnesota, USA. E-mail: {soheil@umn.edu}.}
\thanks{C. Suh is with the School of Electrical Engineering, Korea Advanced Institute of Science and Technology, South Korea. E-mail: {chsuh@kaist.ac.kr}.}%
}

\maketitle

\begin{abstract}
We investigate the role of a relay in multiple access channels (MACs) with bursty user traffic, where intermittent data traffic restricts the users to bursty transmissions. As our main result, we characterize the degrees of freedom (DoF) region of a $K$-user \emph{bursty} multi-input multi-output (MIMO) Gaussian MAC with a relay, where Bernoulli random states are introduced to govern bursty user transmissions. To that end, we extend the noisy network coding scheme to achieve the cut-set bound. Our main contribution is in exploring the role of a relay from various perspectives. First, we show that a relay can provide a DoF gain in bursty channels, unlike in conventional non-bursty channels. Interestingly, we find that the relaying gain can scale with additional antennas at the relay to some extent. Moreover, observing that a relay can help achieve collision-free performances, we establish the necessary and sufficient condition for attaining collision-free DoF. Lastly, we consider scenarios in which some physical perturbation shared around the users may generate data traffic simultaneously, causing transmission patterns across them to be correlated. We demonstrate that for most cases in such scenarios, the relaying gain is greater when the users' transmission patterns are more correlated, hence when more severe collisions take place. Our results have practical implications in various scenarios of wireless networks such as device-to-device systems and random media access control protocols.
\end{abstract}


%
\IEEEpeerreviewmaketitle

\section{Introduction}

Relays have been considered unable to provide degrees-of-freedom (DoF) gains in standard information-theoretic channels in which transmitters are assumed to send signals at all times~\cite{jafar:it09}. From these results, it may seem reasonable to think of relays as not playing much of a role in improving the performance of most channels. The purpose of this work is to show that it is yet premature to conclude so. In practice, transmitters in most wireless networks do not send signals all the time, unlike in the standard information-theoretic models. Rather, they exhibit \emph{bursty} natures. Intermittent data traffic makes the availability of data for transmission limited at transmitters, as opposed to the conventional assumption in the standard models. This leads to bursty signal transmissions. Investigating such practical networks, which we call \emph{bursty} networks hereby, a recent study has found that a relay can offer a DoF gain in bursty interference channels (ICs) and also help achieving interference-free DoF performances~\cite{kim:isit15}. Likewise, subsequently in this work, we demonstrate that a relay in bursty multiple access channels (MACs) can also provide much benefits playing interesting roles.

\subsection{Example and Key Idea}
\begin{figure}[!t]
\centering
\includegraphics[width=0.6\columnwidth]{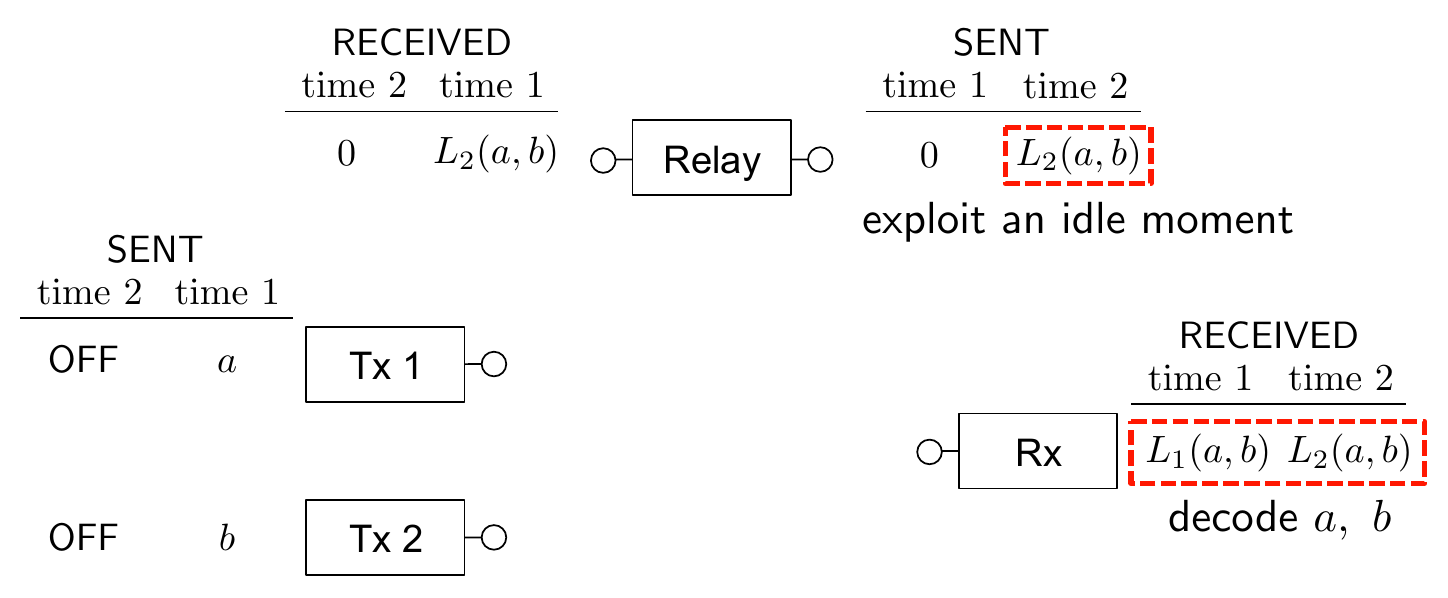}
\caption{An achievable scheme in the two-user single-antenna setting. The relay exploits an idle moment of the transmitters to deliver a useful symbol to the receiver. This relay operation helps resolving a collision.}
\label{fig:scheme}
\end{figure}

\begin{figure}[!t]
\centering
\includegraphics[width=0.45\columnwidth]{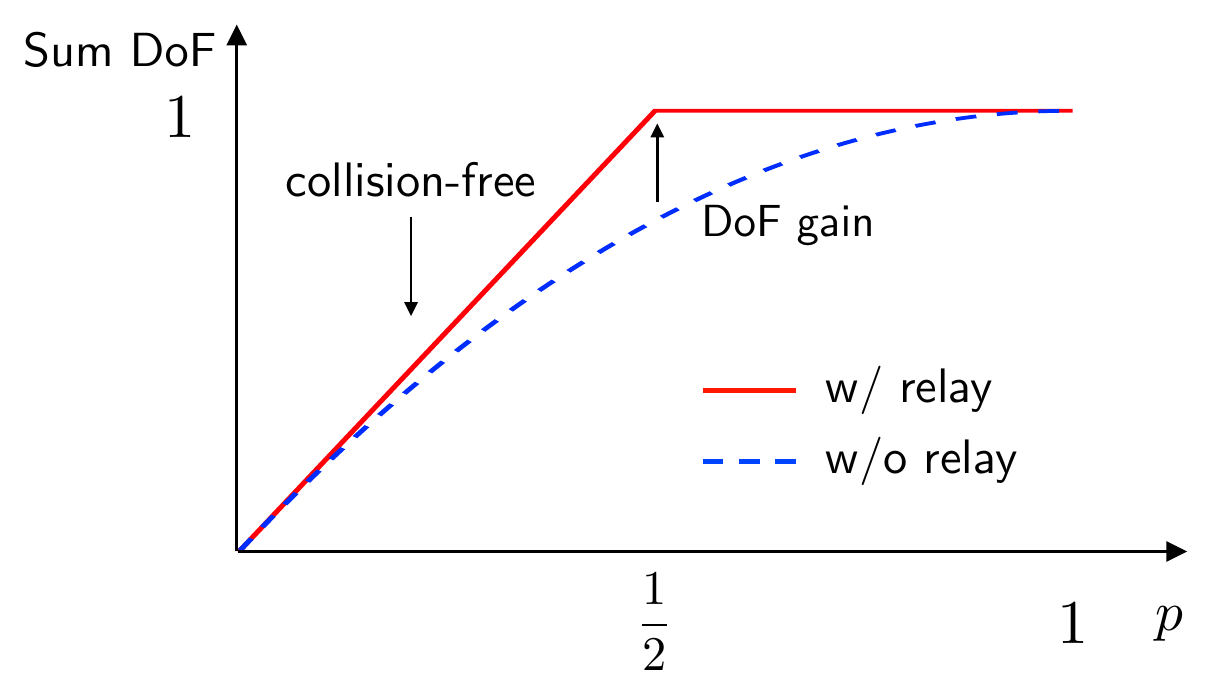}
\caption{The sum DoF in the two-user single-antenna setting with and without a relay. We observe a DoF gain at all levels of burstiness represented by $p$. Interestingly, we also observe a collision-free DoF performance in low data traffic regimes $p < \frac{1}{2}$.}
\label{fig:singleplot}
\end{figure}

Let us first study a simple example to see how employing a relay can be beneficial in bursty MACs. 

\begin{example}
\label{ex:motivating}
Consider a two-user single-antenna setting in which each transmitter becomes active independently with probability $p$. Fig.~\ref{fig:scheme} illustrates how the presence of a relay can help resolving a collision that occurs when both transmitters are active at the same time:
\begin{itemize}
\item Time 1: Both transmitters are active. The receiver gets a linear sum of two symbols. It cannot decode any of them. Meanwhile, the relay gets another linear sum of the symbols. 
\item Time 2: Both transmitters are inactive. The relay forwards its past received linear sum. With the two linear sums, the receiver now can decode the two symbols, resolving the past collision.
\end{itemize}
\end{example}

We see one simple idea at play: The relay can exploit idle moments of the transmitters. It receives symbols that can be useful for the receiver while both transmitters are active, and forwards the symbols to the receiver while they are inactive. This helps resolving collisions. One can verify that the scheme achieves the sum DoF of $\min (2p, 1)$ extending the idea, and also prove its optimality using the cut-set bound. Fig.~\ref{fig:singleplot} compares the sum DoF with a relay and that without a relay. We can observe a DoF gain at all levels of bursty data traffic ($p \neq 0, 1$). More interestingly, we can see that both transmitters achieve the individual DoF of $p$, which is in effect a collision-free DoF performance in low data traffic regimes ($p < \frac{1}{2}$). This result in the two-user single-antenna setting makes us wonder what potential benefits a relay will be able to offer in more general settings: With more users involved, how will the gains from the assistance of a relay vary? With additional antennas installed at the relay, will the gains also increase? In this work, we set out to answer these questions.

\subsection{Main Contribution}
As our main result, extending the key idea shown in Example~\ref{ex:motivating} that a relay can exploit bursty natures of the transmitters, we characterize the DoF region of a $K$-user bursty multi-input multi-output (MIMO) Gaussian MAC with a relay. Our main contribution in this work is exploring the role of a relay in the bursty network of our interest in depth from a practical standpoint, which we anticipate to hold true in other bursty networks as well. To highlight some of the notable findings, first, we demonstrate that a relay can offer a DoF gain in bursty MACs. Given that relays have been known unable to offer a DoF gain in the conventional non-bursty channels, this finding calls for further research on the role of relays in practical wireless networks where transmitters may behave in a bursty fashion. Also, we establish the necessary and sufficient condition for attaining collision-free DoF. Exploring scalability aspects, we find that to some extent the relaying gain can scale as additional antennas are installed at the relay. Moreover, focusing on the scalability with the number of users, we discover that the relaying gain can grow (and soon converge) as the number of involved users increases. Lastly, we examine relaying gains in scenarios where some physical perturbation shared around the users may generate data traffic for them simultaneously, and cause correlated transmission patterns among them. We demonstrate that in most cases the relaying gain is greater when the users exhibit more correlated transmission patterns. This shows that employing a relay can be more beneficial when collisions among the users are likely to be more severe.

\subsection{Practical Implications}
Our results well emphasize advantages of employing a relay in practical wireless systems where multiple sources deliver data to a single destination in a bursty manner. One advantage is to improve performance in emerging networks, such as a device-to-device system in which mobile devices directly exchange data with little help of base stations. When multiple devices convey bursts of data to a device in such systems, what our results suggest is that, the assistance of a relay can be useful to achieve higher data throughput since the relay can provide DoF gains scalable with additional relay antennas, and even better, enable collision-free communication. Another advantage is to simplify random access protocols to reduce control signaling overhead. When multiple users wish to deliver data to a common destination, some protocols are needed to manage signal collisions for reliable data delivery. According to our results, a relay can take a burden off the users when it comes to coping with such collisions. As Example~\ref{ex:motivating} illustrates, the relay can resolve collisions on behalf of the users. Hence, the users are allowed to send signals at random intervals without making extra effort such as retransmissions of collided signals. We discuss more on the practical implications in Section~\ref{sec:implications}.


\subsection{Related Work} 
Extensive research has been done on relay networks since the introduction of the relay channel (RC), in which one source wishes to deliver a message to one destination with the help of a relay, by van der Meulen~\cite{Meulen:it71}. In their seminal paper~\cite{Cover:it79}, Cover and El Gamal laid the foundations for studying relay networks by developing most general coding strategies for the RC, such as decode-and-forward, partial decode-and-forward and compress-and-forward. Among subsequent works that extended the coding schemes to seek capacity for variants of the RC, the most related to our interest is the work due to Kramer-Gastpar-Gupta~\cite{kramer:it05} in which a multiple access relay channel (MARC) is investigated. They developed an achievable scheme by extending decode-and-forward strategies for the RC to a MARC where two sources wish to deliver independent messages to one destination with a relay's aid, and evaluated achievable rates of the scheme in wireless channels with fading. The works due to Lim et al.~\cite{noisy:it11} and Avestimehr-Diggavi-Tse~\cite{salman:it11}, which independently developed coding schemes for noisy relay networks, are most relevant to our results in this paper. We extended the noisy network coding scheme for multimessage multicast networks developed in~\cite{noisy:it11} to characterize the DoF region of the channel of our interest.

In terms of relaying operations, the aforementioned works mainly consider full-duplex and strictly causal relays. Most of the theories developed for full-duplex relays were shown extendable to cases where relays are half-duplex through discussions in~\cite{Kramer:allerton04} which model half-duplex relays by imposing constraints such as fractions of time allowed for relays to be in either reception or transmission mode. Also, causal relaying strategies including instantaneous relaying as well as non-causal relaying strategies have been explored in \cite{ElGamal:it07}.

As we noted earlier, our results have practical implications in random access protocols such as well-known ALOHA~\cite{Abramson:ALOHA}, CSMA~\cite{kleinrock:comm75}, and their extensions. In all such protocols, when users share a common communication medium, they should take part in avoiding and/or recovering from collisions. Our results imply that introducing relays can be effective ways of taking a burden off the users in coping with such collisions, hence simplifying random access protocols. 

Some work has been done on bursty interference networks~\cite{khude:isit09, wang:spawc13, wang:isit13, wang:isit14, kim:isit15}. Khude-Prabhakaran-Viswanath first studied a two-user bursty IC and characterized the capacity region of a deterministic model \cite{khude:isit09}. Wang-Diggavi examined a two-user bursty IC with and without feedback and characterized the capacity region of a bursty Gaussian IC to within a bounded gap \cite{wang:spawc13}. Extending the results in \cite{wang:spawc13}, Kim-Wang-Suh investigated a bursty IC with feedback in the presence of a relay and partially established the necessary and sufficient condition for attaining interference-free DoF \cite{kim:isit15}.
 
To the best of our knowledge, there has been little work done on multiuser networks where correlations across the users' transmission patterns are taken into account. Such correlations across bursty user transmissions have been out of question, as almost all related past works have considered conventional non-bursty channels where transmitters send signals at all times.

\section{Problem Formulation}
\label{sec:model}
\begin{figure}[!t]
\centering
\includegraphics[width=0.6\columnwidth]{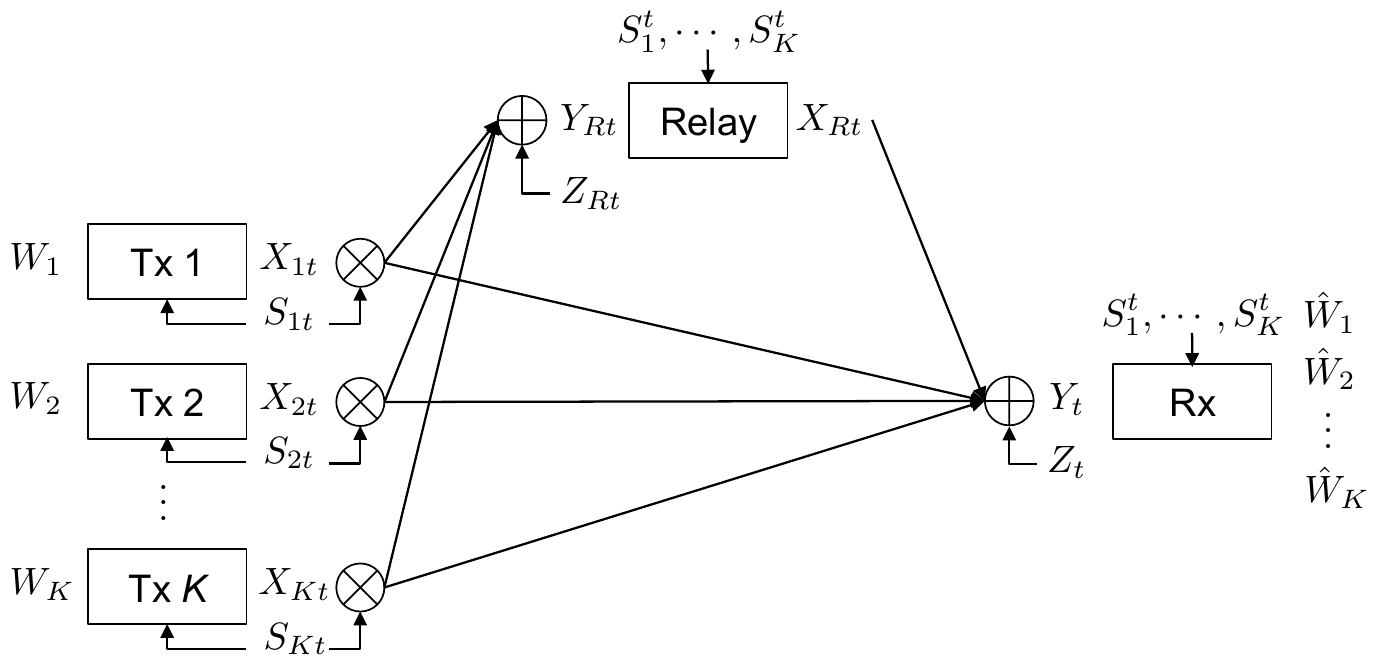}
\caption{$K$-user bursty MIMO Gaussian MAC with a relay.}
\label{fig:model}
\end{figure}


Fig.~\ref{fig:model} describes the $K$-user bursty MIMO Gaussian MAC with a relay. Transmitter $k$ has $M_k$ antennas, $\forall k = 1, 2, \dots, K$, and the receiver and the relay have $N$ and $L$ antennas respectively. Transmitter $k$ wishes to reliably deliver a message $W_k$ to the receiver.

Let $X_{kt} \in \mathbb{C}^{M_k}$ be transmitter $k$'s encoded signal and $X_{Rt} \in \mathbb{C}^L$ be the relay's encoded signal at time $t$. The encoded signals are power-constrained: $\mathbb{E}[|X_{kt}^2|] \leq P$ and $\mathbb{E}[|X_{Rt}^2|] \leq P$. Traffic states $S_{kt}$ are assumed to follow $\mathsf{Bern}(p)$ to govern bursty transmissions over time. Joint distribution\footnote{We let joint distributions capture the users' transmission patterns that stem from correlations across the data availabilities, while for simplicity we assume marginal distributions to be identical.} $\mathsf{P}(S_{1}, S_{2}, \dots, S_{k})$ captures the bursty transmission patterns of the users. 
Unlike the transmitters, the relay is not subject to bursty transmissions.

It is worthwhile to note the rationale behind our modeling of bursty transmissions. This work is motivated by sensor networks in which transmitters are simple devices, thus it is natural to consider them to be less capable; they can neither have a large buffer to store ample data for later use, nor employ advanced scheduling schemes. Hence, intermittent data traffic forces the users to send signals whenever they have available data to deliver, leading to bursty transmissions. In contrast, a relay can be equipped with richer capabilities such as a large buffer and channel state sensing; it can store sufficient past received signals and use them later according to channel states for better assistance. Hence, unlike the transmitters, the relay can send signals at all times. Also, as we consider intermittent data availabilities to be a primary cause of bursty transmissions, it might be more practically relevant to model such burstiness in higher layers of the communication protocol stack. However, to greatly simplify the model while capturing the bursty nature to some extent, we incorporate multiplicative random states into the physical channel.


Additive noise terms $Z_{t}$ at the receiver and $Z_{Rt}$ at the relay are assumed to be independent of each other and of transmit signals, distributed according to $\mathcal{CN}(0,\mathbf{I}_N)$ and $\mathcal{CN}(0,\mathbf{I}_L)$, and i.i.d. over time. Let $Y_{t} \in \mathbb{C}^N$ be the received signal at the receiver and $Y_{Rt} \in \mathbb{C}^L$ be the received signal at the relay at time $t$. Then,
\begin{align*}
Y_{t} = \sum_{k} \mathbf{H}_{k} S_{kt} {X}_{kt} + \mathbf{H}_{R} X_{Rt} + Z_{t}, \quad Y_{Rt} = \sum_{k} \mathbf{H}_{Rk} S_{kt} {X}_{kt}  + Z_{Rt}.
\vspace*{-0.5in}
\end{align*}
The matrices $\mathbf{H}_{k}$ and $\mathbf{H}_{Rk}$ describe the time-invariant channels from transmitter $k$ to the receiver and to the relay, respectively. The matrix $\mathbf{H}_{R}$ describes the time-invariant channel from the relay to the receiver. All channel matrices are assumed to be full rank.

We assume current traffic states are available at the receiver and the relay.
Each transmitter knows its own current traffic state, as it has access to the availability of its own data for transmission.


Transmitter $k$ encodes its signal at time $t$ based on its own message and its own current traffic state; we assume uncoordinated traffic states across all transmitters, which means each transmitter has access to its own traffic state only. The relay encodes its signal at time $t$ based on its past received signals, and both past and current traffic states of all transmitters. Then,
\begin{align*}
X_{kt} = f_{kt}(W_k, S_{k}^t), \quad X_{Rt} = f_{Rt}(Y_R^{t-1}, \boldsymbol{S}^t).
\end{align*}
For notational convenience, we let $\boldsymbol{S}_t := (S_{1t}, S_{2t}, \dots, S_{Kt})$ and $\boldsymbol{S}^{t}$ stand for the sequence up to $t$.

Finally, we define the DoF region $\mathcal{D} := \{ ( d_1, d_2, \dots, d_K ) : \exists (R_1, R_2, \dots, R_K) \in \mathcal{C}(P) \text{ and } d_k = \lim_{P \to \infty} \frac{R_k}{\log P} \}$,
where $\mathcal{C}(P)$ is the capacity region with power constraint $P$ and $R_k$ is the data rate of transmitter $k$.

\section{Main Results} \label{sec:mainresults}

As our main result, we characterize the DoF region of the $K$-user bursty MIMO Gaussian MAC with a relay.
{\it \begin{theorem} \label{thm:dofregion}
The DoF region of the $K$-user bursty MIMO Gaussian MAC with a relay is characterized as follows.
\begin{align}
\sum_{k \in \mathcal{K}} d_k \leq \min
\left[
\sum\limits_{\mathcal{A \subseteq \mathcal{K}}} {\sf P}(\mathcal{A}) \min \left( \sum\limits_{i \in \mathcal{A}} M_i, N+L \right),
\sum\limits_{\mathcal{A \subseteq \mathcal{K}}} {\sf P}(\mathcal{A}) \min \left( \sum\limits_{i \in \mathcal{A}} M_i + L, N \right)
\right], \label{eq:dofregion}
\end{align}
where $\mathcal{K} \subseteq \{1,2,\dots,K\}$, $\mathcal{A}$ is a set that indicates which transmitters are active, and ${\sf P}(\mathcal{A})$ is a joint distribution that describes the probabilities of the transmitters indicated by $\mathcal{A}$ being active.
\end{theorem}}
\begin{IEEEproof}
Here we provide an outline of the proof. See Appendix~\ref{app:thmproof} for the details. To obtain an outer bound, we directly follow the standard cut-set arguments~\cite{gamalkim:nit}. To obtain a matching inner bound, we extend noisy network coding for multimessage multicast networks~\cite{noisy:it11}. Introducing traffic states information is the key to the extension. We let the transmitters and the relay not make use of any traffic states information although they have access to (part of) it, whereas we let the receiver do so. This is equivalent to the case where information of traffic states is available only at the receiver. Hence, we treat the received signal and the traffic states (i.e., $(Y_t, \boldsymbol{S}_t)$) as the output of the channel at time $t$. Also, we apply amplify-and-forward strategies without compression at the relay. Then, we evaluate the extended achievable scheme using independent Gaussian input distributions, to get an achievable DoF region that matches with the cut-set bound.
\end{IEEEproof}

For ease of presentation, in this paper we mainly consider a symmetric $K$-user bursty MAC with a relay in which all transmitters have the same number of antennas: $M_i = M, ~\forall i = 1,2\dots,K$. Thus, we state the following corollary.
{\it \begin{corollary} \label{cor:dofregion}
The DoF region of the symmetric $K$-user bursty MIMO Gaussian MAC with a relay in which all transmitters have the same number of antennas is characterized as follows.
\begin{align}
\sum_{k \in \mathcal{K}} d_k \leq \min
\left[
\sum\limits_{\mathcal{A \subseteq \mathcal{K}}} {\sf P}(\mathcal{A}) \min \left( |\mathcal{A}|M, N+L \right),
\sum\limits_{\mathcal{A \subseteq \mathcal{K}}} {\sf P}(\mathcal{A}) \min \left( |\mathcal{A}|M + L, N \right)
\right].
\end{align}
\end{corollary}}
\begin{IEEEproof}
Setting $M_i = M, ~\forall i = 1,2\dots,K$ in Theorem~\ref{thm:dofregion} completes the proof.
\end{IEEEproof}

To investigate the number of additional DoF attained with the assistance of a relay, we focus on additive sum DoF gains: the difference between the sum DoF with a relay and that without a relay. 
{\it \begin{corollary} \label{cor:gendofgain}
The additive sum DoF gain attained by employing a relay in the symmetric $K$-user bursty MIMO Gaussian MAC is expressed as follows. 
\begin{align}
\Delta \mathsf{DoF}
= & \min
\left[ \begin{array}{l}
\sum\limits_{\mathcal{A} \in \Omega} \mathsf{P}(\mathcal{A}) \min (|\mathcal{A}|M, N+L), 
\sum\limits_{\mathcal{A} \in \Omega} \mathsf{P}(\mathcal{A}) \min (|\mathcal{A}|M+L, N)
\end{array} \right] - \sum_{\mathcal{A} \in \Omega} \mathsf{P}(\mathcal{A}) \min (|\mathcal{A}|M, N), \label{eq:sumdofgain}\end{align}
where $\Omega$ is the set that includes all subsets of $\{1, 2, \dots, K\}$.
\end{corollary}}
\begin{IEEEproof}
Setting $L = 0$ in Corollary~\ref{cor:dofregion} gives rise to the sum DoF region without a relay. Subtracting it from the sum DoF with a relay in Corollary~\ref{cor:dofregion} completes the proof.
\end{IEEEproof}

Although it might be a comprehensive approach to explore how the gains attained from employing a relay change with varying levels of correlations across the users' data availabilities (equivalently, their transmission patterns), we consider two extreme ends of the spectrum for simplicity. On the one end, the users' data are available in a fully dependent manner, thus all transmitters are either active or inactive at the same time. On the other end, the users' data are available in an independent manner, thus a transmitter becomes active regardless of the others.
{\it \begin{corollary} \label{cor:dofgains}
The additive sum DoF gain attained by employing a relay in the symmetric $K$-user bursty MIMO Gaussian MAC with fully dependent and independent transmissions are expressed as follows.
\begin{align}
\Delta \mathsf{DoF_{dep}} = & \min
\left[~ p \min (KM-N, L), (1-p) \min (L, N) ~\right], \label{eq:depgain} \\
\Delta \mathsf{DoF_{ind}} = & \min
\left[~ \sum\limits_{i=\lfloor \frac{N}{M} \rfloor+1}^{K} \mathsf{B}_{K}(i) \min (iM-N, L), \sum\limits_{i=0}^{\lfloor \frac{N}{M} \rfloor} \mathsf{B}_{K}(i) \min (L, N-iM) ~\right], \label{eq:indgain}
\end{align}
where ${\sf B}_{K}(i) := \binom{K}{i} p^i (1-p)^{K-i}$.
\end{corollary}}
\begin{IEEEproof}
See Appendix~\ref{app:dofgainsproof}.
\end{IEEEproof}

\section{Collision-free DoF and Gain Scalabilities}
\label{sec:maindiscussion}

We now begin in-depth discussions on a variety of topics: collision-free DoF performances and scalability aspects of the relaying gains.

\subsection{Collision-free DoF Performances} \label{sec:collisionfree}



Collision-free communication may be the most desirable benefit one hopes to get from the assistance of a relay. No user needs to step back for the sake of others. As we can see in Example~\ref{ex:motivating}, in low data traffic regimes, the aid of a relay can help achieving collision-free DoF performances. When the data traffic level is low, idle moments of the transmitters take place more often than collisions occur. During the idle moments, the relay can send useful symbols to the receiver that help resolving the collisions, leading to the desirable performances. When the data traffic level is high, on the other hand, the relay finds fewer opportunities to send and the number of decodable symbols flowing into the receiver is limited. Hence, saturated DoF performances are achieved instead.

In this section, we establish the necessary and sufficient condition for attaining collision-free DoF. For simplicity, we consider the case where the users send bursty signals in an independent manner. The proof of necessity is in Appendix~\ref{app:necsuf}. The proof of sufficiency follows by Corollary~\ref{cor:dofregion}.
{\it \begin{corollary} \label{cor:necsuf}
The necessary and sufficient condition for attaining collision-free DoF for $p \in (0, p^*)$, where\footnote{We assume $KM > N$ throughout this paper. Otherwise, there is no point in discussing relaying gains, as the receiver is able to decode all transmit symbols instantaneously.} $p^* \leq \frac{N}{KM}$, in the $K$-user bursty MIMO Gaussian MAC with a relay is expressed as
\begin{align} \label{eq:necsuf}
KM \leq N+L.
\end{align}
\end{corollary}}

Condition~(\ref{eq:necsuf}) includes an obvious case in which the number of receive antennas is greater than or equal to the total number of transmit antennas ($KM \leq N$). In this case, the receiver can decode all transmit symbols instantaneously without any help of a relay, even when all transmitters are active at the same time. We can achieve the collision-free DoF of $KMp$ without a relay, and hence it is of no use.

In the other case ($KM > N$), condition~(\ref{eq:necsuf}) says that we need a relay to achieve collision-free DoF performances and that the relay should have at least $KM - N$ antennas. This is a condition that intuitively comes to mind. When all transmitters are active at the same time, the number of linear combinations of the transmit symbols the relay gets should be at least as large as the receiver additionally needs for successful decoding of all transmit symbols.


\begin{figure}[!t]
\centering
\includegraphics[width=\columnwidth]{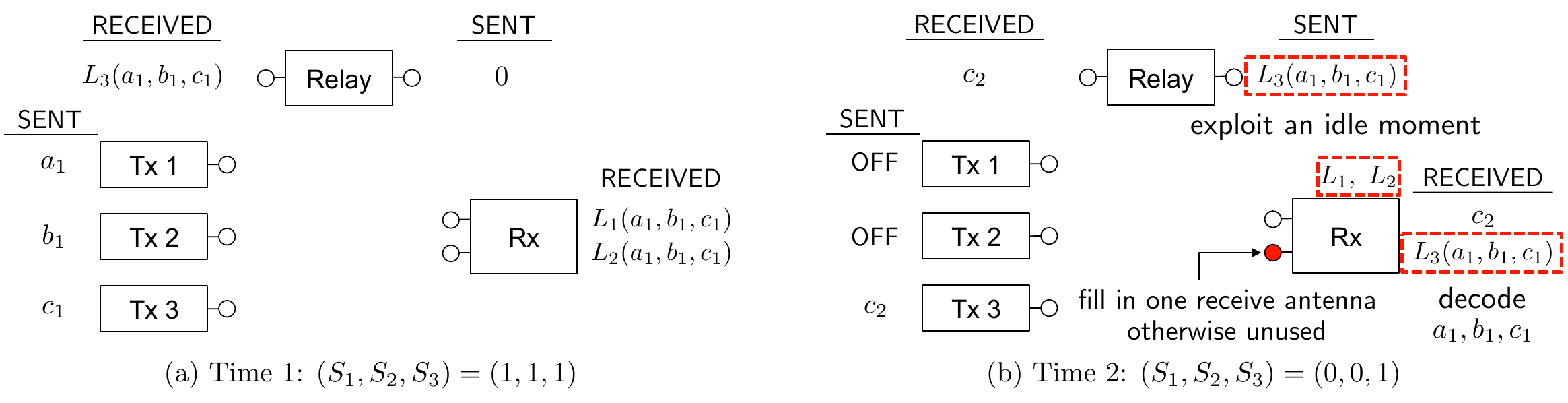}
\caption{The relay fills in otherwise unused receive antennas with symbols that help resolving collisions. In low data traffic regimes, this happens frequently enough, leading to collision-free DoF performances.}
\label{fig:free}
\end{figure}


What remains is to determine what operation of the relay makes it possible to achieve collision-free DoF performances. In proving sufficiency of~(\ref{eq:necsuf}), for attaining collision-free DoF, we extend the noisy network coding scheme~\cite{noisy:it11}. When extending the scheme, we let the relay receive-and-forward without compression. Fig.~\ref{fig:free} illustrates the relay operation in a simple network of $(K,M,N,L) = (3,1,2,1)$: when only a few transmitters (or none) are active, the relay fills in unused antennas of the receiver with useful symbols, i.e., the symbols that help resolving past collisions. Again apparent is the motivating key idea of this work: a relay exploiting idle moments of the transmitters. In low data traffic regimes, it is more likely that only a few transmitters are active. Then, compared to the rate of occurrence at which collisions take place, the relay more frequently finds opportunities to deliver symbols to the receiver that are intended for resolving past collisions. This leads to collision-free DoF performances.

{\bf Remark:}
There is an interesting difference to note between the relay operations in bursty MACs and bursty ICs. A recent study on a two-user bursty MIMO Gaussian IC with a relay, focusing on interference-free DoF performances, develops a scheme in which the relay \emph{cooperates} with active transmitters~\cite{kim:isit15}. From this cooperation, they neutralize interference in the air, which coincides with the idea in~\cite{mohajer:it11}. This suggests that more sophisticated operations of the relay may be required to achieve optimality in other multiuser bursty networks.


\subsection{Scalability with the Number of Users} \label{sec:userscalability}

In this section, we aim to investigate how the DoF gain attained from employing a relay varies as the number of users grows. In doing so, we consider scenarios in which dependencies across the users' intermittent data availabilities may cause correlated transmission patterns of the users. We explore two representative cases: one concerns fully dependent transmission patterns and the other concerns independent ones. That is, we focus on the additive sum DoF gains $\Delta {\sf DoF_{dep}}$ and $\Delta {\sf DoF_{ind}}$ given in Corollary~\ref{cor:dofgains}, and their scalability in terms of the number of users involved.


As in Section~\ref{sec:collisionfree}, we continue to assume the best-case scenario in which sufficiently many antennas are installed at the relay (more precisely $L \geq (KM-N)^+$), so that we can examine the case where the relay can help achieving the maximum (collision-free) DoF. Considering a motivating example of device-to-device systems, which consists of mostly mobile users with limited antenna availabilities, we assume $M=N=1$. To simplify matters further, making the number of users the only variable, we focus on one specific point in low data traffic regimes: $p^* = \frac{1}{K}$. A rationale behind this choice is that at this level of data traffic, both $\Delta {\sf DoF_{dep}}$ and $\Delta {\sf DoF_{ind}}$ have their peak values (See Appendix~\ref{sec:peakproof}), which we denote by $\Delta {\sf DoF_{dep}^{peak}}$ and $\Delta {\sf DoF_{ind}^{peak}}$ respectively, and formally define as follows:
\begin{align*}
\Delta {\sf DoF_{dep}^{peak}} := \max_{p} \Delta {\sf DoF_{dep}}, \quad \Delta {\sf DoF_{ind}^{peak}} := \max_{p} \Delta {\sf DoF_{ind}}.
\end{align*}
It means that each peak value represents the maximal relaying gain for each case of user transmission patterns, given a fixed number of users.

\begin{figure}[!t]
\centering
\includegraphics[width=0.45\columnwidth]{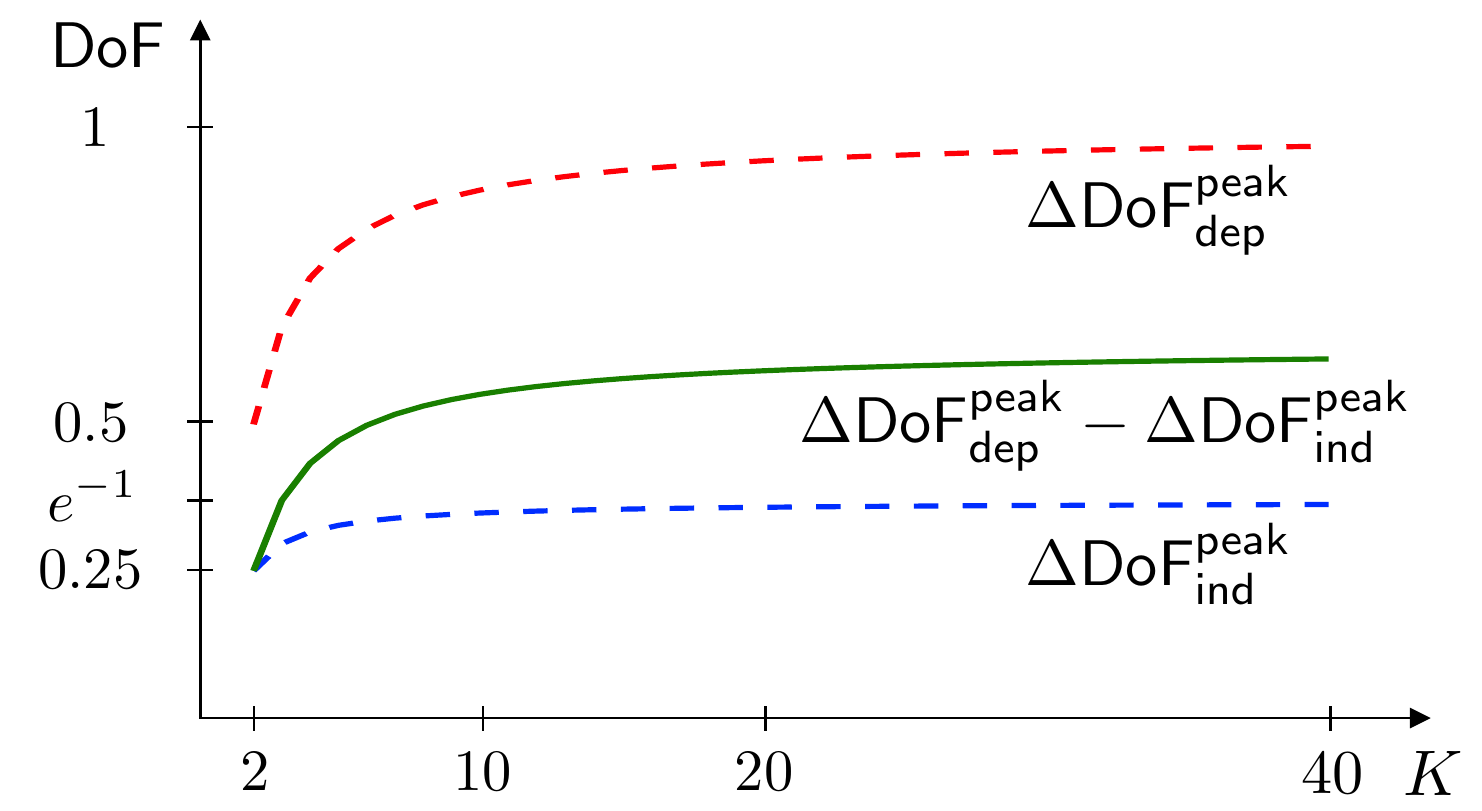}
\caption{The growing peak DoF gains with both dependent and independent data availabilities, and the expanding gap between them. Antenna settings: $M=N=1$ and $L \geq K-1$.}
\label{fig:wrt_k}
\end{figure}

We obtain the following peak relaying gains as a function of $K$ from Corollary~\ref{cor:dofgains} under the assumptions (See Appendix~\ref{sec:peakproof}):
\begin{align*}
\Delta \mathsf{DoF_{dep}^{peak}} = 1-\frac{1}{K}, \quad \Delta \mathsf{DoF_{ind}^{peak}} = \Big( 1-\frac{1}{K} \Big)^K.
\end{align*}
One can easily verify that both peak gains grow as $K$ increases, and converge to 1 and $\frac{1}{e}$, respectively. This finding shows that the beneficial role of a relay does not deteriorate as the number of users increases. To the contrary, the presence of a relay is more advantageous with more users regardless of dependencies across the users' data availabilities.
Interestingly, comparing the two peak gains, we can see that the difference between both peak gains ($\Delta \mathsf{DoF_{dep}^{peak}} - \Delta \mathsf{DoF_{ind}^{peak}}$) grows as $K$ increases (See Appendix~\ref{sec:peakproof}).
This finding shows that the difference in the amounts of additional DoF obtained with the assistance of a relay expands as the number of users increases.

Based on the discussion above, one may conclude that in the best-case scenario we consider, employing a relay in bursty MACs is (1) more advantageous when more users are involved, and (2) more beneficial with dependencies across the users' data availabilities. In Section~\ref{sec:relaygains}, we compare the relaying gains for the two cases of distinct user transmission patterns in further detail to see if the conclusion still holds true for other scenarios as well.



\subsection{Scalability with the Number of Relay Antennas} \label{sec:scalability}


So far, we have optimistically assumed sufficiently many relay antennas for the sake of exploring the best-case scenario in which employing a relay can be most beneficial. Now, let us limit the number of relay antennas. Having an economic perspective in mind, it is natural for one to weigh the benefit that comes from installing one extra relay antenna in comparison with the associated cost, thus to be interested in incremental relaying gains. When one intends to employ a relay into real-world wireless systems and wishes to benefit from it, what is designable with a certain extent of freedom would be the number of relay antennas; the number of users involved in the systems and the antenna configuration of the users would be given within a plausible range. Hence, exploring another scalability aspect of the DoF gains in this section, we pay attention to incremental relaying gains in terms of the number of relay antennas, for a fixed number of users. 



As in Section~\ref{sec:userscalability}, we assume $M = N = 1$. Also, we consider the case where there are four users, i.e., $K=4$.
As before, we consider two representative cases: one concerns fully dependent transmission patterns and the other concerns independent ones. In high traffic regimes, there is little scalability to observe due to saturated DoF at the receiving users. Thus, we mainly focus on low traffic regimes.


\begin{figure}[!t]
\centering
\includegraphics[width=0.9\columnwidth]{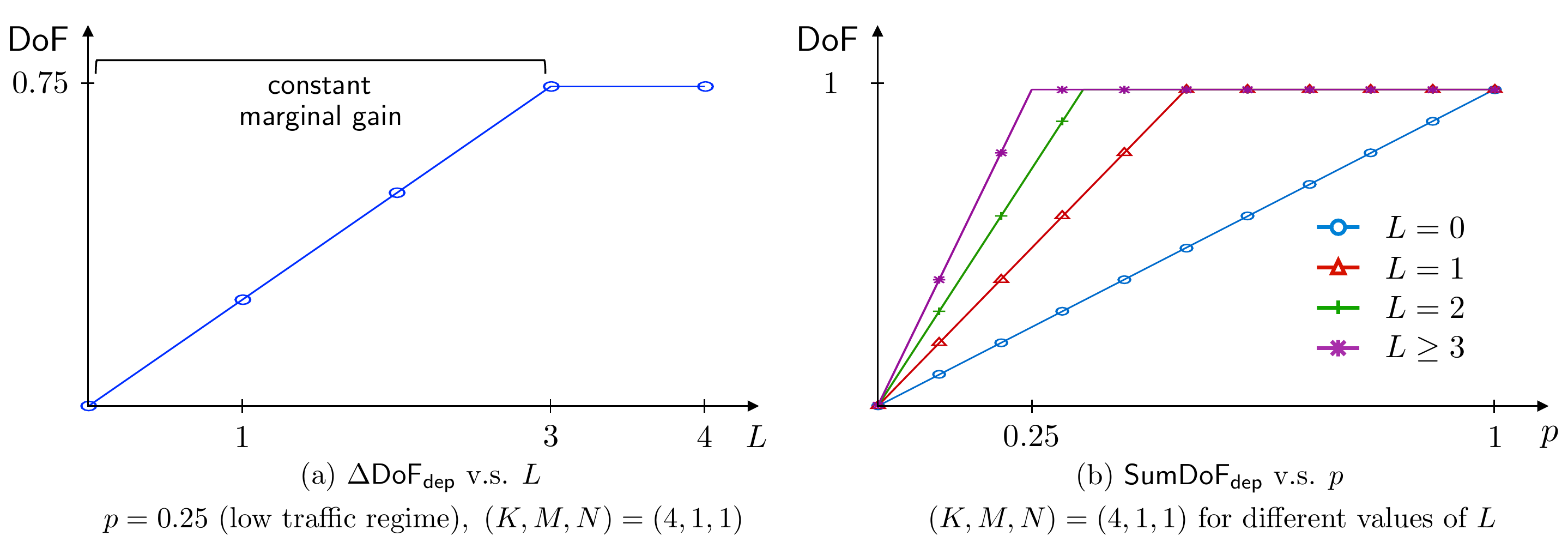}
\caption{$K = 4$ and $M=N=1$. (a): $\Delta {\sf DoF_{dep}}$ v.s. $L$ in low traffic regimes $p = 0.25$. We can observe a constant incremental gain with the number of relay antennas, and when the number exceeds a threshold, we observe no more gain with an extra relay antenna. (b): relaying gains for all levels of burstiness. With very high traffic, little scalability is observed; one extra relay antenna is enough to achieve the maximum DoF.}
\label{fig:marginal_dep}
\end{figure}

Fig.~\ref{fig:marginal_dep}(a) illustrates how the DoF gain in the presence of a relay varies as the number of relay antennas changes, when the users show fully dependent transmission patterns: $\Delta {\sf DoF_{dep}}$ v.s. $L$. We can observe a \emph{constant} incremental gain to some extent, namely until $L = 3$ which is equal to $KM-N$. Once the number of relay antennas exceeds it, the relaying gain vanishes. This shows that, when the users exhibit fully dependent transmission patterns, the first few relay antennas are helpful, each with equal worth, and soon an extra at the relay does not help at all. Hence, from an economic perspective, it is best to install antennas at the relay as many as $KM-N$.




\begin{figure}[!t]
\centering
\includegraphics[width=0.9\columnwidth]{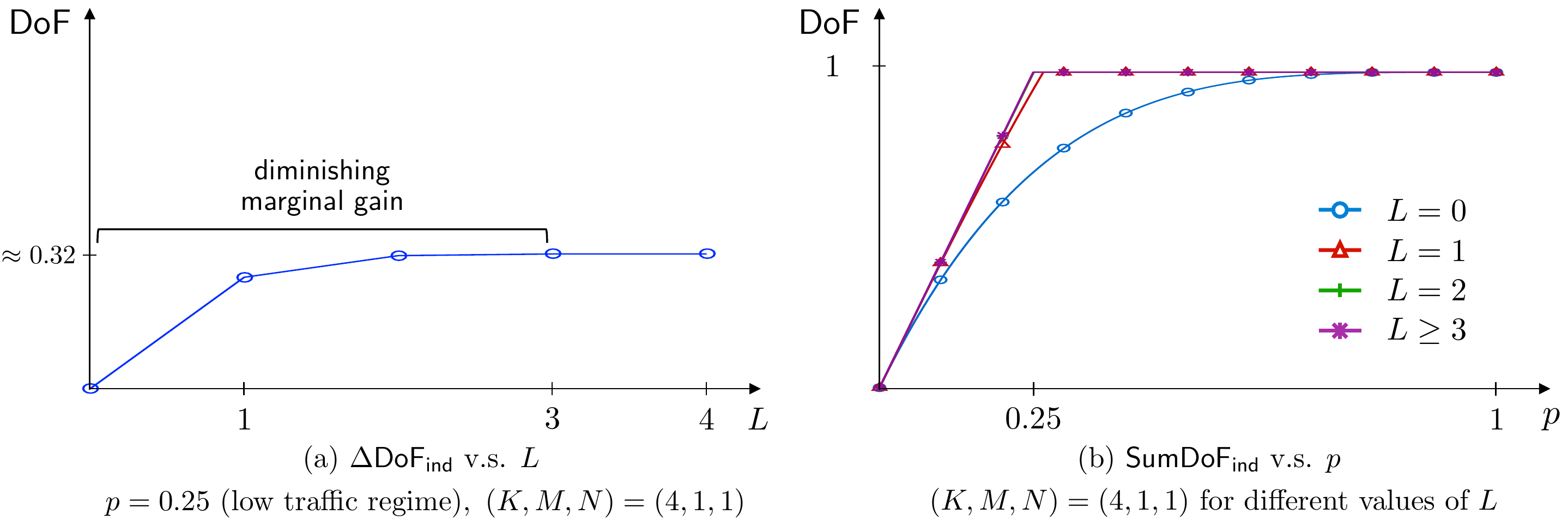}
\caption{$K = 4$ and $M=N=1$. (a): $\Delta {\sf DoF_{ind}}$ v.s. $L$ in low traffic regimes $p = 0.25$. We can observe a diminishing incremental gain with the number of relay antennas, and when the number exceeds a threshold, we observe no more gain with an extra relay antenna. (b): relaying gains for all levels of burstiness. With very high traffic, little scalability is observed; one extra relay antenna is enough to achieve the maximum DoF.}
\label{fig:marginal_ind}
\end{figure}

Fig.~\ref{fig:marginal_ind}(a) illustrates how the DoF gain in the presence of a relay varies as the number of relay antennas changes, when the users show independent transmission patterns: $\Delta {\sf DoF_{ind}}$ v.s. $L$. We can observe a \emph{diminishing} incremental gain to some extent, namely until $L = 3$ which is equal to $KM-N$. Once the number of relay antennas exceeds it, the relaying gain vanishes. This shows that, when the users exhibit independent transmission patterns, the first few antennas are helpful, each with decreasing worth, and soon an extra at the relay does not help at all. Hence, from an economic perspective, if the cost of installing an extra antenna at the relay is big, so the benefit of it should be properly weighed, it is better to install a small number of them, since more will not result in much bigger benefits.


Figs.~\ref{fig:marginal_dep}(b) and \ref{fig:marginal_ind}(b) illustrate that in high traffic regimes the presence of a relay still helps. However, due to saturated DoF at the receiving users, at high traffic levels, the value of an extra relay antenna decreases much faster; one is worth the same as two or more in this example (particularly in Fig.~\ref{fig:marginal_ind}(b), one relay antenna gives almost the same benefit, so we barely notice the difference), hence little scalability is observed.

\begin{figure}[!t]
\centering
\includegraphics[width=0.5\columnwidth]{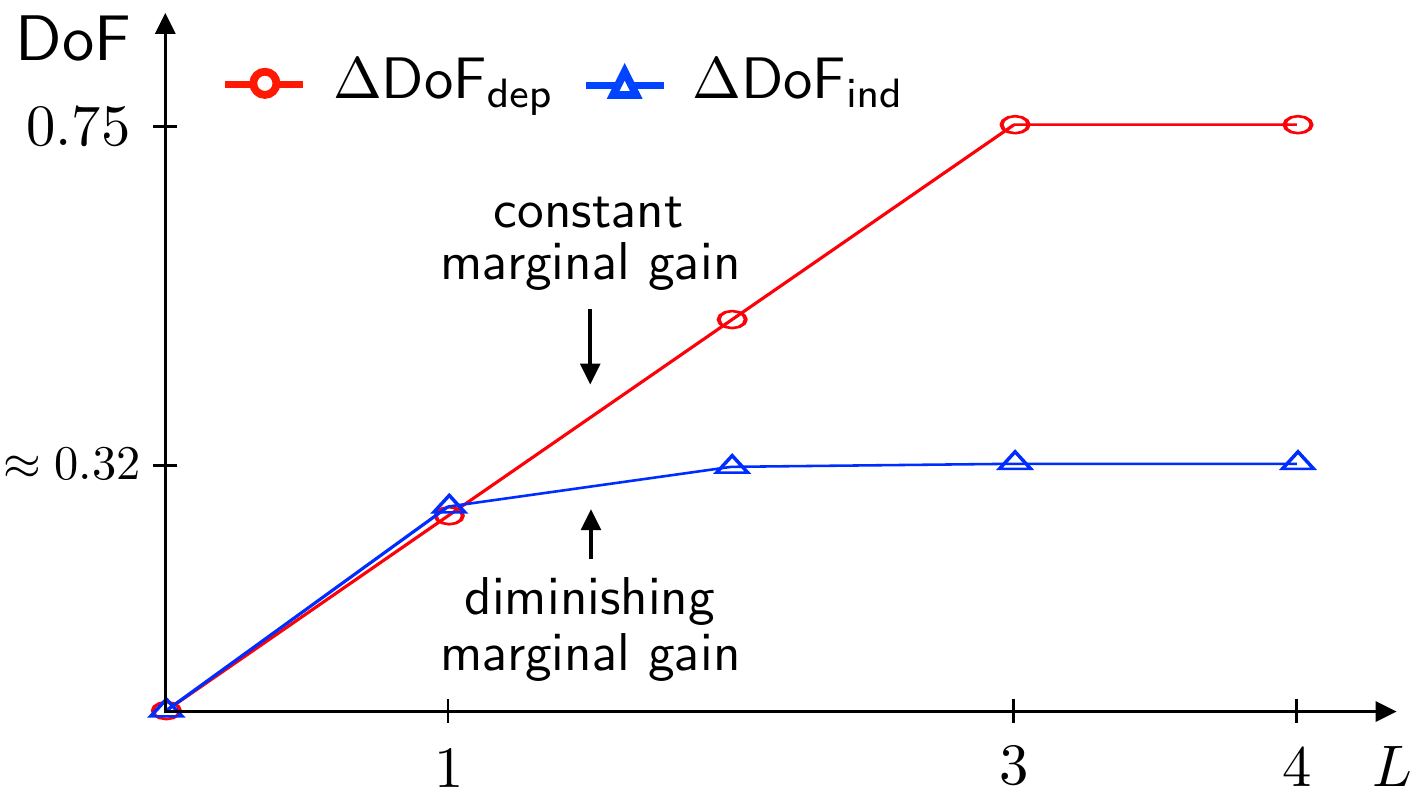}
\caption{$\Delta {\sf DoF_{dep}}$ and $\Delta {\sf DoF_{ind}}$ v.s. $L$ in low traffic regimes $p = 0.25$. $K = 4$ and $M = N = 1$.}
\label{fig:marginal_both}
\end{figure}

Fig.~\ref{fig:marginal_both} compares the scalabilities of both $\Delta {\sf DoF_{dep}}$ and $\Delta {\sf DoF_{ind}}$ with respect to $L$. One thing we can infer from this figure is that in realistic scenarios where the users' transmission patterns are uncorrelated or moderately correlated (not fully), an extra relay antenna will be worth the most when it is the first to be installed, due to the diminishing incremental gain. A few other antennas to be installed will be worth less. Another thing to note is that as Section~\ref{subsec:N L KM-N} pointed out that a relay could provide a greater gain with uncorrelated user transmissions in low traffic regimes with antenna configurations $N \leq L \ll KM-N$, we can observe that the relaying gain with uncorrelated user transmissions is (slightly) greater than that with correlated ones when $L = 1$, confirming the point. In the other cases when $L \geq 2$, we see greater relaying gains with correlated user transmissions.


{\bf Remark:}
The limited scalability can be explained with an analogy. In bursty MACs, there is one receiver to which all transmitters wish to deliver their message; they are sharing one pie. One transmitter sending at higher rates necessarily means the other transmitters sending at lower rates. Employing a relay is shown to be beneficial, but not significantly. The relay may help the transmitters consume the pie better, but after all, it cannot increase the size of the pie no matter how well it operates.

In bursty interference channels (ICs), however, there is a different story to tell. Recently, it is shown that a relay can offer a DoF gain in bursty ICs that can scale \emph{linearly} with additional relay antennas~\cite{kim:isit15}. In the IC case, in contrast with the MAC case, each transmitter wishes to deliver their message to its own intended receiver; they are not sharing one pie. One transmitter sending at higher rates could mean the other transmitters sending at lower rates, but it does not result from exclusively consuming the pies of the others. Rather, it results from hindering the others from having theirs.
The relay can help the transmitters consume their own pie only, so that each consumes its own, not hindered by the others.


\section{Relaying Gains: When Transmission Patterns Across Users May Be Correlated} \label{sec:relaygains}
Consider a sensor network in which multiple sensors gather temperature measurements and convey them to a central hub which performs some control task. Alternatively, consider a safety network in which nearby vehicles equipped with sensors to detect a possible risk share information to prevent accidents from happening. In both example networks, some physical perturbation around objects in close proximity may cause the objects to collect and exchange bursts of data simultaneously. Such dependencies across the users' intermittent data availabilities may cause correlated transmission patterns across the users to take place.

In this section, we set out to investigate how the DoF gains in the presence of a relay vary according to correlations across the users' transmission patterns. We consider the two representative cases as in the previous sections: the case of fully dependent user transmissions and the case of independent user transmissions. We compare the relaying gains in the two cases for various antenna configurations, as the role of a relay and its functionality in networks may vary depending on them.

\subsection{$L \geq \max \left( KM-N, N \right)$: Relays with enough antennas}
The condition implies that the relay is able to get additional signals that the receiver needs to resolve the worst-case collisions that occur when all transmitters are active, and also able to forward the maximum number of signals that the receiver can get at a time. For this case, the relay can help achieving the maximum DoF for a given bursty MAC: $\min(pKM, N)$, that is collision-free DoF in the low traffic regime and saturated DoF in the high traffic regime. Intuitively, the relay receives additional signals when too many transmitters are active, and later forwards them when only a few are active, to achieve the maximum DoF.

\begin{figure}[!t]
\centering
\includegraphics[width=0.45\columnwidth]{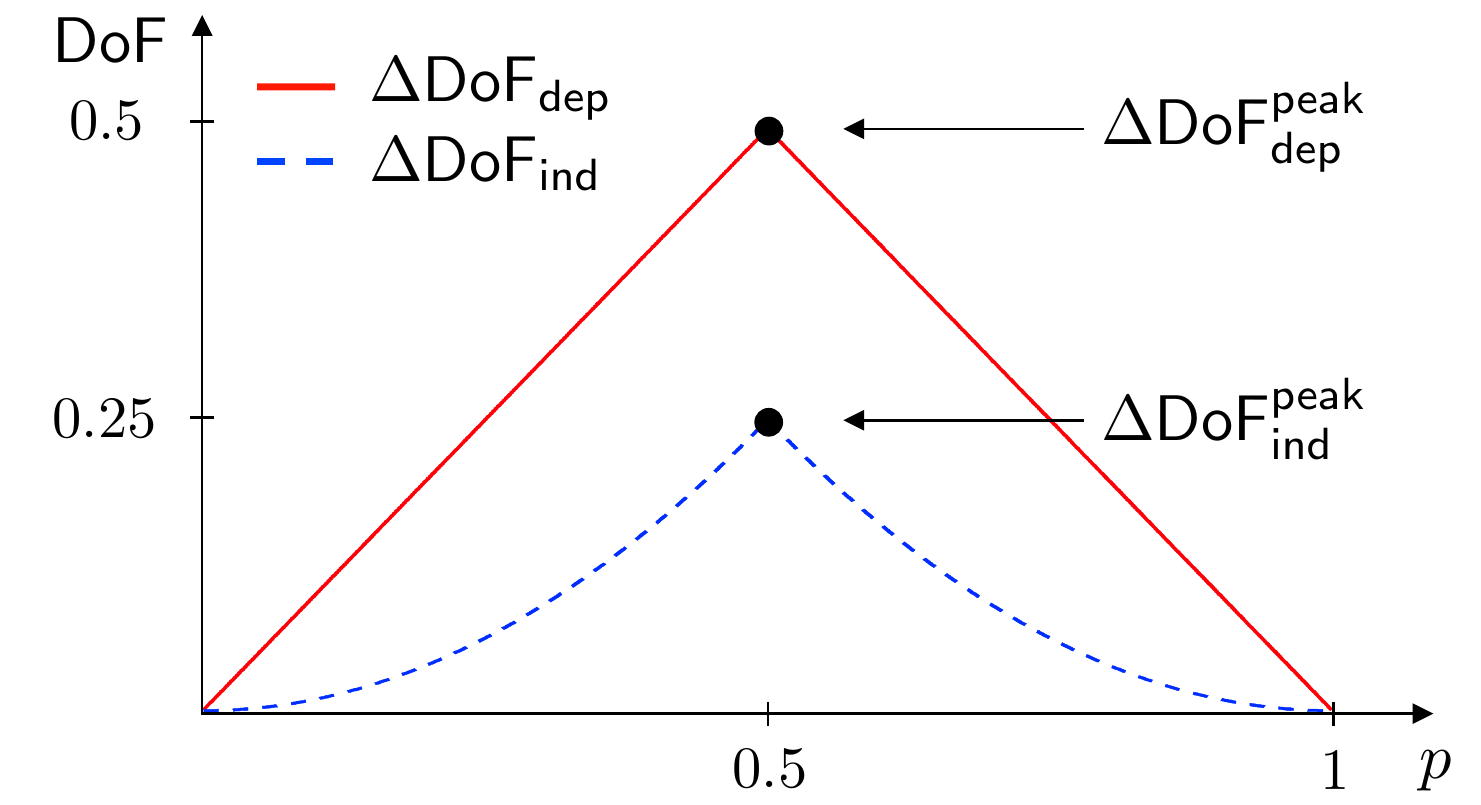}
\caption{A relay provides greater DoF gains when the users' data availabilities are dependent. Antenna setting: $(K,M,N,L) = (2,1,1,1)$.}
\label{fig:wrt_p}
\end{figure}

We find that the gain obtained from the assistance of a relay is greater with dependent data availabilities than that with independent data availabilities. This finding is illustrated in Fig.~\ref{fig:wrt_p}, and in mathematical terms,
\begin{align}
\Delta \mathsf{DoF_{dep}} > \Delta \mathsf{DoF_{ind}}, ~ p \in (0,1). \label{depgreaterthanind}
\end{align}
See Appendix~\ref{sec:gainproof} for the proof. For this case, in the presence of a relay, the same sum DoF $\min(pKM, N)$ can be achieved regardless of dependencies across the users' data availabilities. Let us see what happens in its absence. With dependent data availabilities, too many symbols are sent simultaneously compared to the number of symbols that the receiver can get at a time. Without a relay, there would be a big DoF loss due to the severe collisions. With independent data availabilities, however, such severe collisions occur less often given the same level of data traffic. Hence, there would be a relatively smaller DoF loss. The absence of a relay costs bursty MACs with dependent data availabilities more, that is, its presence is more beneficial, offering greater DoF gains.

We note that this finding is valid for the case where the relay has sufficiently many antennas to help achieving the maximum DoF. In the rest of this section, we examine other cases where the relay does not have enough number of antennas to help achieving the maximum DoF, and hence offers limited DoF gains.


\subsection{$KM-N \leq L < N$}
\label{subsec:KM-N L N}
A relay offers DoF gains operating in two different modes. It receives additional symbols in reception mode, and forwards them in transmission mode. For this case, particularly when $L \ll N$, the relay may reveal its drawback in transmission mode. In this mode, the relay forwards additional symbols to the receiver, to provide DoF gains by utilizing the receive antennas otherwise unused. In that sense, $L \ll N$ indicates its limited capability to utilize all receive antennas.

\begin{figure}[!t]
\centering
\includegraphics[width=0.7\columnwidth]{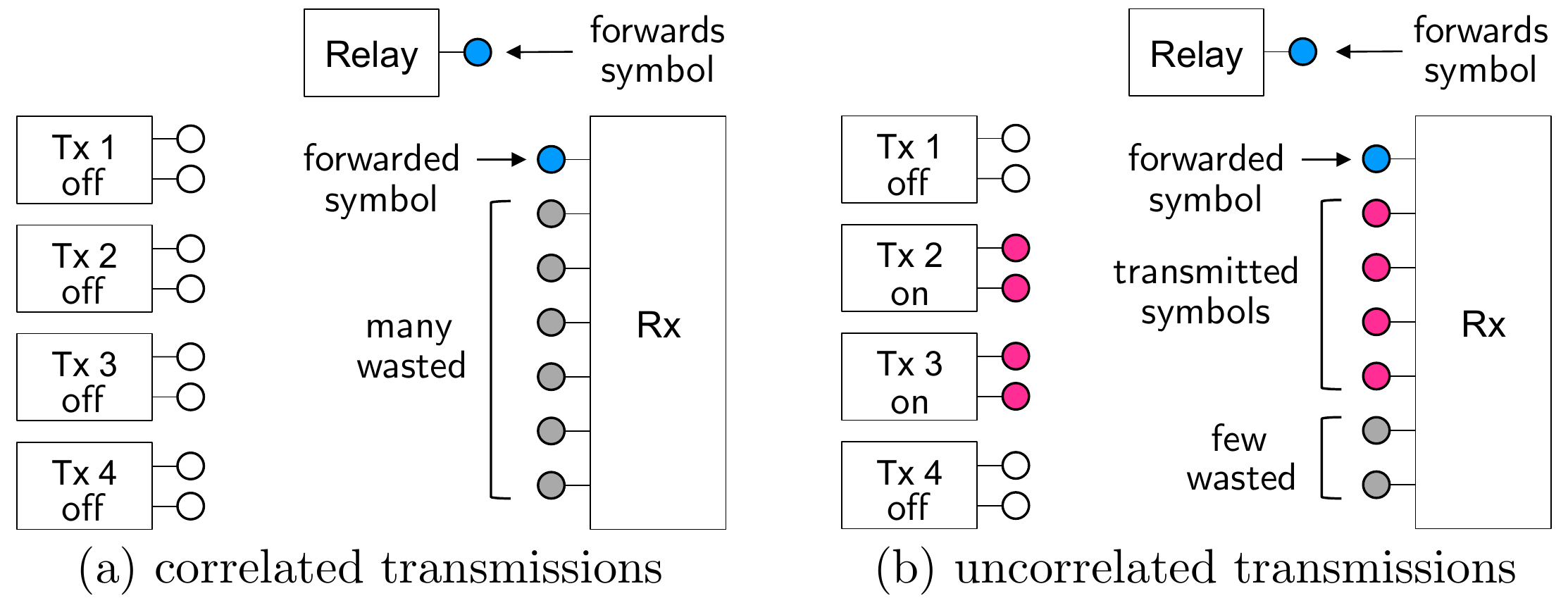}
\caption{Antenna setting $(K,M,N,L) = (4,2,7,1)$: $KM-N \leq L < N$. Limited capability of the relay in transmission mode stands out with dependent user data availabilities. See Fig.~\ref{fig:limited}(a).}
\label{fig:tx_phase}
\end{figure}

The limitation stands out with dependent data availabilities, and makes the relay less beneficial with such dependencies. With dependent data availabilities, all transmitters are either active or inactive. When they are inactive, the relay operates in transmission mode to provide a DoF gain. However, the relay can utilize a very small fraction of receive antennas due to its drawback $L \ll N$, leaving a large fraction of them wasted (Fig.~\ref{fig:tx_phase}(a)). With independent data availabilities, on the other hand, these undesired incidents do not occur as often, given the same data traffic level (Fig.~\ref{fig:tx_phase}(b)).

Fig.~\ref{fig:limited}(a) illustrates the relay providing greater DoF gains with independent data availabilities in high $p$ regimes only. This is because in those regimes, the relay is guaranteed to receive enough additional symbols from the transmitters, hence the number of additional symbols it can forward in transmission mode is what determines the amount of DoF gain it offers. In short, the limitation in transmission mode ($L \ll N$), which appears only in high $p$ regimes, affects the relay's capability more adversely with dependent data availabilities, hence makes the relay less beneficial with correlated user transmissions.





\subsection{$N \leq L < KM-N$}
\label{subsec:N L KM-N}

\begin{figure}[!t]
\centering
\includegraphics[width=0.7\columnwidth]{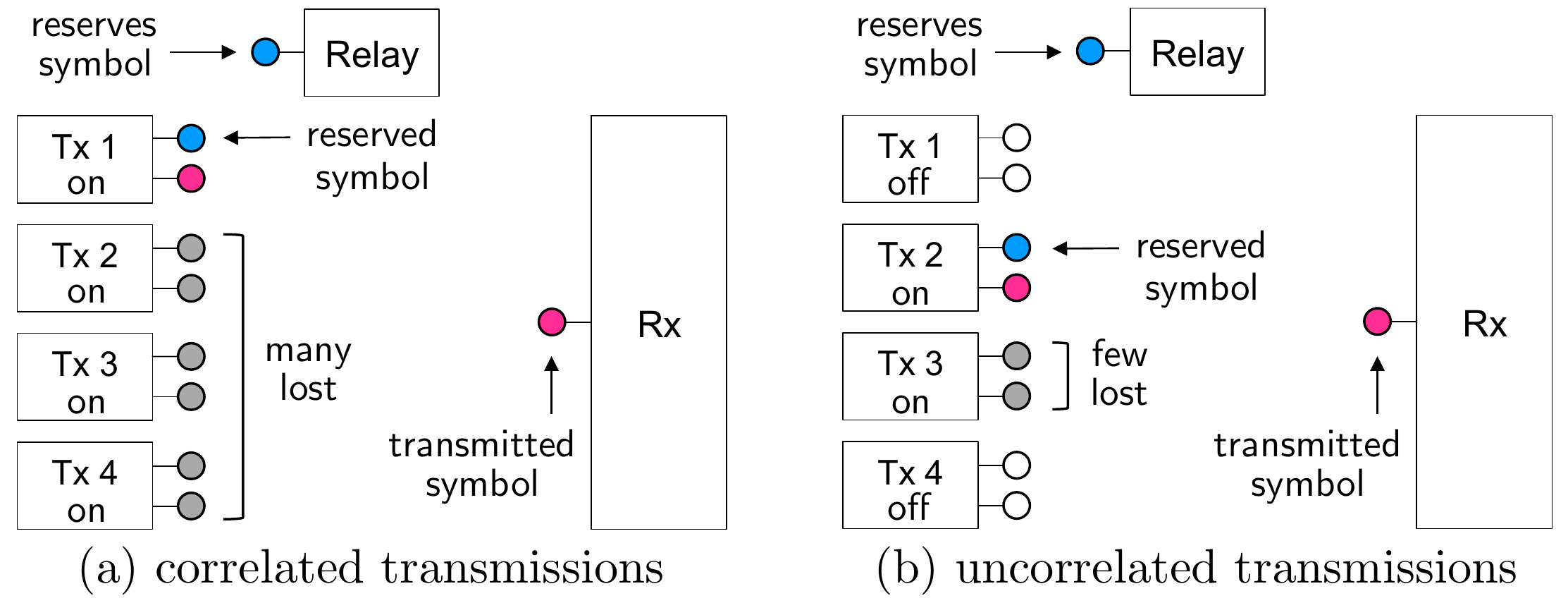}
\caption{Antenna setting $(K,M,N,L) = (4,2,1,1)$: $N \leq L < KM-N$. Limited capability of the relay in reception mode stands out with dependent user data availabilities. See Fig.~\ref{fig:limited}(b).}
\label{fig:rx_phase}
\end{figure}

\begin{figure}[!t]
\centering
\includegraphics[width=0.7\columnwidth]{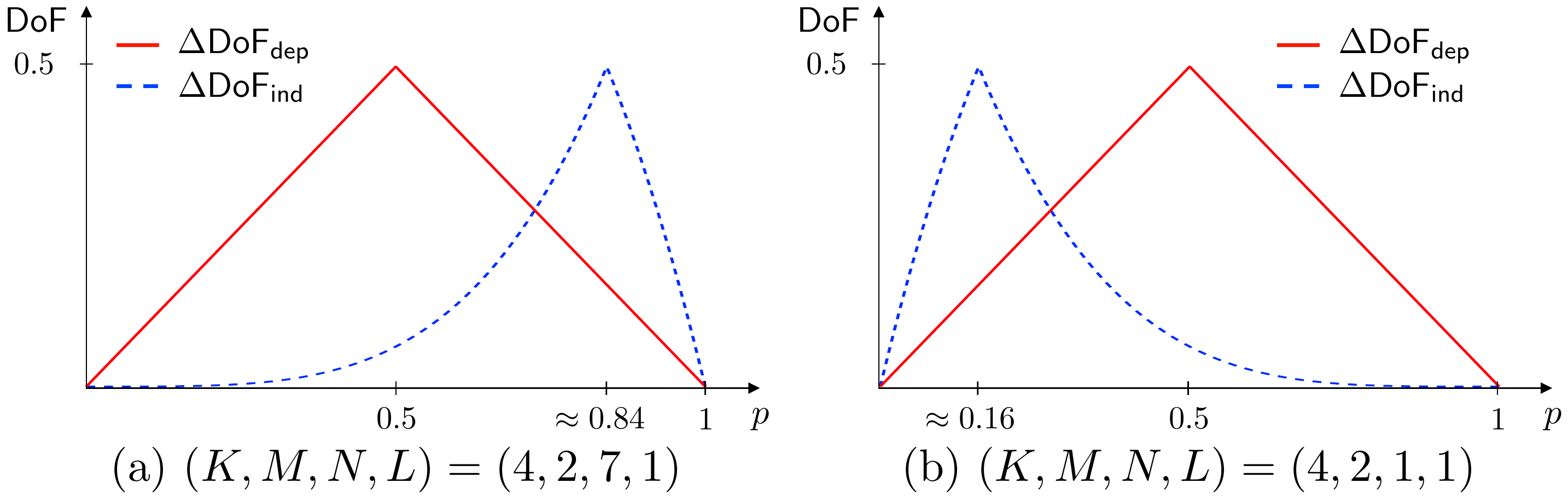}
\caption{DoF gains with independent user data availabilities are greater than those with dependent user data availabilities: when $KM - N \leq L < N$, with high traffic (left) and when $N \leq L < KM-N$, with low traffic (right).}
\label{fig:limited}
\end{figure}


For this case, particularly when $L \ll KM-N$, the relay may reveal its drawback in reception mode. In this mode, the relay receives and reserves the additional symbols from active transmitters otherwise lost, to provide DoF gains by forwarding them later to the receiver. In that sense, $L \ll KM-N$ indicates its limited capability to reserve all surplus symbols.

The limitation stands out with dependent data availabilities, and makes the relay less beneficial with such dependencies. With dependent data availabilities, when all transmitters are active, the relay operates in reception mode to provide a DoF gain. However, the relay can reserve a very small fraction of surplus symbols due to its drawback $L \ll KM-N$, leaving a large fraction of them lost (Fig.~\ref{fig:rx_phase}(a)). With independent data availabilities, on the other hand, these undesired incidents do not occur as often, given the same data traffic level (Fig.~\ref{fig:rx_phase}(b)).

Fig.~\ref{fig:limited}(b) illustrates the relay providing greater DoF gains with independent data availabilities in low $p$ regimes only. This is because in those regimes, the relay is guaranteed to find enough idle moments of the transmitters to forward additional symbols to the receiver, hence the number of additional symbols it can receive in reception mode is what determines the amount of DoF gain it offers. In short, the limitation in reception mode ($L \ll KM-N$), which appears only in low $p$ regimes, affects the relay's capability more adversely with dependent data availabilities, hence makes the relay less beneficial with correlated user transmissions.



We have not examined the case $L < \min \left( KM-N, N \right)$. The condition implies that the relay's limited capabilities may be revealed in either reception or transmission mode, leading to its worse performance with dependent data availabilities. Regimes in which it underperforms depend on the antenna settings: if $L \ll N$ then it has limitations in transmission mode hence underperforms in high $p$ regimes, and if $L \ll KM-N$ then in low $p$ regimes.

\section{Practical Implications} \label{sec:implications}
The bursty model in this work can be naturally translated into many practical wireless systems. In this section, we discuss implications of our results in such systems with some examples. 

\subsection{Connection with Device-to-device Systems}
\label{subsec:iot}
As mobile communication capabilities have continued to be improved, device-to-device systems are receiving attention. As introduced in the beginning of this paper, bursty MACs can be a suitable model that constitute the systems. A simple example of device-to-device systems can be a sensor network consisting of a central hub with multiple sensors that computes the average room temperature: the sensors located at various places deliver measurements to the hub, and the hub computes the average. Another can be a safety network consisting of vehicles with sensors in which when the vehicles spot one vehicle facing a possible risk, they provide the vehicle in danger the information to prevent the accident in advance.

Emerging device-to-device systems in their initial deployment stage are less likely to comprise of objects with many antennas installed. The objects, which previously would have been deficient of communication capabilities, will share bursts of small amounts of data with elementary communication to carry out a task that is better done with collective data from many sources. In that sense, it is reasonable to assume that the number of users $K$ is large, the numbers of transmit and receive antennas $M$ and $N$ are small, and the level of data traffic $p$ is low. This scenario well fits with the bursty MACs we have investigated, and here we can ask a natural question: if we were to employ a relay to help all objects deliver data at their best (optimistically, without any collisions among them), how many antennas should we install at the relay and how should the relay operate?


Our results say that by employing a relay with at least $KM - N$ antennas that performs a simple receive-and-forward operation, we can make all objects trying to deliver bursts of data to a common destination convey their data effectively without collisions. Not only does the relay increase overall data throughput, it also has practical benefits in reducing control signaling overhead. When data traffic is sufficiently low, the relay eases the need of the devices to coordinate their transmissions to avoid possible performance degradation. It also eases the need of exchanging acknowledge signals among them to let each other know the receptions of previously sent signals.

\subsection{Connection with Random Media Access Control Protocols}
\label{subsec:protocol}
When we formulated our problem, we considered intermittent data traffic as a primary source of bursty transmissions. They can, however, result from a random media access control protocol with which multiple transmitters sharing a common medium comply. Now, the transmitters may send signals in a bursty manner, not because data to transfer is intermittently available, but because they intend to avoid collisions. Although the source of burstiness is different, our bursty model captures this scenario well.

Let us consider a wireless system with a simple protocol. $K$ transmitters with $M$ transmit antennas wish to send signals to a receiver with $N$ receive antennas. There is a relay with enough antennas from which this wireless system can benefit. We assume $M < N$ as the receiver may have more antennas than the multiple transmitters that are connected to it, and $KM > N$ as it may have not so many to accommodate all transmitters at once. To avoid collisions that possibly degrade overall performance, each transmitter sends signals according to a protocol: sending signals with \mbox{probability $p$} and making such decisions independently over time. In this case, \mbox{probability $p$} no longer represents bursty data traffic as in our original model. Rather, it is now a design parameter of the system. A natural question to ask is: how do we choose $p$ to achieve the best performance?

Our results say that by choosing $p = \frac{N}{KM}$, we can achieve the best performance, with no one transmitter lowering its performance for the sake of the others. This threshold probability makes sense, since the relay schedules bursty transmissions of all transmitters and lets them equally share the receiver. The scheduling role of the relay relaxes the complication of the protocol. There is no need of feeding back some information from the receiver to the transmitters to manage collisions.

\section{Conclusion}
\label{sec:concl}
We characterized the DoF region of the $K$-user bursty MIMO Gaussian MAC with a relay. To that end, we extended the noisy network coding scheme to achieve the DoF cut-set bound. From the characterization, we found that a relay can provide a DoF gain in bursty MACs, and also established the necessary and sufficient condition for attaining collision-free DoF. Interestingly, we demonstrated that the relaying gain can scale with additional relay antennas to some extent. Examining the gain in further detail, we showed that the relaying gain is in most cases greater when the users exhibit more correlated transmissions across each other and when more users are involved, except for some cases in which the relay is equipped with a very small number of antennas. Our results have practical implications of employing a relay into wireless systems where multiple sources wish to deliver data to a single destination in a bursty manner, in terms of better performance and media access control protocols design.



%

\appendices
\section{Proof of Theorem~\ref{thm:dofregion}} \label{app:thmproof}

First, we state an outer bound on the capacity region, which directly follows by the standard cut-set arguments, omitting the proof:

\begin{lemma} \label{lem:outer}
Any achievable rate region $(R_1, R_2, \dots, R_K)$ of the $K$-user bursty MIMO Gaussian MAC with a relay must satisfy the inequalities such that
\begin{align*}
\sum_{k \in \mathcal{K}} R_k \leq \left[ \begin{array}{l} I \left( X(\mathcal{K}); Y, Y_R | \boldsymbol{S}, X(\mathcal{K}^c), X_R \right), I \left( X(\mathcal{K}), X_R ; Y | \boldsymbol{S}, X(\mathcal{K}^c) \right) \end{array} \right],
\end{align*}
for all distributions $F(X_1, X_2, \dots, X_K, X_R)$ such that $\mathbb{E}[X_k^2] \leq P$ and $\mathbb{E}[X_R^2] \leq P$, where $\mathcal{K} \subseteq \{1, 2, \dots, K\}$ and $\boldsymbol{S} := (S_1, S_2, \dots, S_K)$.
\end{lemma}
To get the DoF outer bound that matches the claimed DoF region (\ref{eq:dofregion}), we evaluate the above cut-set bound with Gaussian distributions with power constraint $P$ that maximize the mutual information terms~\cite{gamalkim:nit}. And we take the limit of $P \rightarrow \infty$ after dividing the evaluated cut-set bound by $\log (P)$. Then, we get the DoF outer bound that matches the claimed DoF region (\ref{eq:dofregion}).

Next, we state an inner bound on the capacity region, which we will prove shortly by extending the noisy network coding scheme~\cite{noisy:it11}:
\begin{lemma} \label{lem:inner}
An achievable rate region of the $K$-user bursty MIMO Gaussian MAC with a relay includes the set of $(R_1, R_2, \dots, R_K)$ such that (without time-sharing)
\begin{align*}
\sum_{k \in \mathcal{K}} R_k < \min \left[ \begin{array}{l}
I( X(\mathcal{K}) ; Y, \hat{Y}_R | \boldsymbol{S}, X(\mathcal{K}^c), X_R ), \\
I( X(\mathcal{K}), X_R ; Y | \boldsymbol{S}, X(\mathcal{K}^c) ) - I(Y_R ; \hat{Y}_R | \boldsymbol{S}, X_1, \dots, X_K, X_R, Y) \end{array} \right]
\end{align*}
for some distribution $\prod_{k=1}^{K} F(x_k)F(x_R)F(\hat{y}_R|y_R, x_R)$ such that $\mathbb{E}[X_k^2] \leq P$ and $\mathbb{E}[X_R^2] \leq P$.
\end{lemma}
To get the DoF inner bound that matches the claimed DoF region~(\ref{eq:dofregion}), we evaluate the achievable rate region with the independent Gaussian distributions with power constraint $P$. And we take the limit of $P \rightarrow \infty$ after dividing the evaluated achievable rate region by $\log (P)$. Then, the rate penalty term, the subtracted mutual information term, vanishes for some choice of $\hat{Y}_R$ in the limit of $P \rightarrow \infty$ as it does not scale with $P$ (we will show this shortly), and we get the DoF inner bound that matches the claimed DoF region~(\ref{eq:dofregion}). Note the similarities of Lemmas~\ref{lem:outer} and \ref{lem:inner}.

Now we begin to prove Lemma~\ref{lem:inner}. For simplicity, we provide a detailed proof for the two-user setting. It is straightforward to extend the proof for the $K$-user setting. By directly applying the noisy network coding scheme~\cite{noisy:it11}, we can get an inner bound. We assume that the transmitters and the relay do not make use of any information of traffic states, although they have access to (part of) it. Hence, transmitter $k$ encodes its signal at \mbox{time $t$} based on its message: $X_{kt} = f_{kt}(W_k)$. The relay encodes its signal at \mbox{time $t$} based on its past received signals: $X_{Rt} = f_{Rt}(Y_R^{t-1})$. On the other hand, the receiver makes use of information of traffic states. Thus, we treat $(Y_t, S_t)$ as the output of the channel at \mbox{time $t$}. We get the following inner bound (without time-sharing):
\begin{align*}
R_1 & < \min \left[ \begin{array}{l} I(X_1 ; (Y, \boldsymbol{S}), \hat{Y}_R | X_2, X_R), \\ I(X_1, X_R ; (Y, \boldsymbol{S}) | X_2) - I (Y_R ; \hat{Y}_R | X_1, X_2, X_R, (Y, \boldsymbol{S})) \end{array} \right], \\ 
R_2 & < \min \left[ \begin{array}{l} I(X_2 ; (Y, \boldsymbol{S}), \hat{Y}_R | X_1, X_R), \\ I(X_2, X_R ; (Y, \boldsymbol{S}) | X_1) - I (Y_R ; \hat{Y}_R | X_1, X_2, X_R, (Y, \boldsymbol{S})) \end{array} \right], \\
R_1 + R_2 & < \min \left[ \begin{array}{l}  I(X_1, X_2 ; (Y, \boldsymbol{S}), \hat{Y}_R | X_R), \\ I(X_1, X_2, X_R ; (Y, \boldsymbol{S})) - I (Y_R ; \hat{Y}_R | X_1, X_2, X_R, (Y, \boldsymbol{S})) \end{array} \right],
\end{align*}
for some distribution $F(x_1)F(x_2)F(x_R)F(\hat{y}_R | y_R, x_R )$ such that $\mathbb{E}[X_1^2] \leq P$, $\mathbb{E}[X_2^2] \leq P$, and $\mathbb{E}[X_R^2] \leq P$.

The traffic states $(S_{1t}, S_{2t})$ are independent of the messages $(W_1, W_2)$ and the noise at the relay $(Z_R^t)$. Also, they are i.i.d. over time. Since $X_{kt} = f_{kt}(W_k)$ and $X_{Rt} = f_{Rt}(Y_R^{t-1})$, the traffic states $(S_{1t}, S_{2t})$ are independent of $(X_{1t}, X_{2t}, X_{Rt})$. Therefore, $I(X_{k \in \mathcal{B}} ; S | X_{k \in \mathcal{B}^c}) = 0$, where $\mathcal{B} \subseteq \{ 1,2,R \}$. Using the chain rule and this equality, we calculate the mutual information terms and get the following inner bound:
\begin{align*}
R_1 & < \min \left[ \begin{array}{l} I(X_1 ; Y, \hat{Y}_R | \boldsymbol{S}, X_2, X_R), \\ I(X_1, X_R ; Y |\boldsymbol{S}, X_2) - I (Y_R ; \hat{Y}_R | \boldsymbol{S}, X_1, X_2, X_R, Y) \end{array} \right], \\ 
R_2 & < \min \left[ \begin{array}{l} I(X_2 ; Y, \hat{Y}_R | \boldsymbol{S}, X_1, X_R), \\ I(X_2, X_R ; Y |\boldsymbol{S}, X_1) - I (Y_R ; \hat{Y}_R | \boldsymbol{S}, X_1, X_2, X_R, Y) \end{array} \right], \\
R_1 + R_2 & < \min \left[ \begin{array}{l}  I(X_1, X_2 ; Y, \hat{Y}_R | \boldsymbol{S}, X_R), \\ I(X_1, X_2, X_R ; Y | \boldsymbol{S}) - I (Y_R ; \hat{Y}_R | \boldsymbol{S}, X_1, X_2, X_R, Y) \end{array} \right],
\end{align*}
for some distribution $F(x_1)F(x_2)F(x_R)F(\hat{y}_R | y_R, x_R )$ such that $\mathbb{E}[X_1^2] \leq P$, $\mathbb{E}[X_2^2] \leq P$, and $\mathbb{E}[X_R^2] \leq P$.

We compute the rate penalty term using almost the same method in~\cite{gamalkim:nit}. We set $\hat{Y}_R = Y_R + \hat{Z}_R$, where $\hat{Z}_R \sim \mathcal{CN}(0, \mathbf{I}_L)$ and is independent of $(\boldsymbol{S}, X_1, X_2, X_R, Y, Y_R)$. We get the following rate penalty:
\begin{align*}
I(Y_R ; \hat{Y}_R | \boldsymbol{S}, X_1, X_2, X_R, Y) & \overset{(a)}{=} h(\hat{Y}_R | \boldsymbol{S}, X_1, X_2, X_R, Y) - h(\hat{Y}_R | \boldsymbol{S}, X_1, X_2, X_R, Y, Y_R) \\
& \overset{(b)}{\leq} h(\hat{Y}_R | \boldsymbol{S}, X_1, X_2, X_R) - h(\hat{Y}_R | \boldsymbol{S}, X_1, X_2, X_R, Y, Y_R) \\
& \overset{(c)}{=} h(Z_R + \hat{Z}_R) - h(\hat{Z}_R) \overset{(d)}{=} L \log  \left( 2 \pi e \cdot 1 \right) - L \log \left( 2 \pi e \cdot \frac{1}{2} \right) = L,
\end{align*}
where $(a)$ follows by the chain rule; $(b)$ follows by the fact that conditioning reduces differential entropy; $(c)$ follows by the fact that $\hat{Z}_R$ is independent of $(\boldsymbol{S}, X_1, X_2, X_R, Y, Y_R)$; $(d)$ follows by the fact that $Z_R \sim \mathcal{CN}(0, \mathbf{I}_L)$ and $\hat{Z}_R \sim \mathcal{CN}(0, \mathbf{I}_L)$ are independent. 


Again note the similarities of Lemmas~\ref{lem:outer} and \ref{lem:inner}. Setting aside the rate penalty terms, the mutual information terms are almost identical except for $Y_R$ and $\hat{Y}_R$. We are mainly concerned with regimes where $P$ is very large. That is, the rate penalty terms, which are shown above not to scale with $P$, will vanish in taking the limit $P \rightarrow \infty$. Also, the noise term that differentiates $Y_R$ and $\hat{Y}_R$ (i.e., $\hat{Z}_R = \hat{Y}_R - Y_R$) will be negligible as its power does not scale with $P$ either. Therefore, we will get the matching DoF inner and outer bounds, and as a result characterize the DoF region of the two-user bursty MIMO Gaussian MAC with a relay. It is straightforward to extend the proof to the $K$-user setting. The exact same line of reasoning presented above holds.

\section{Proof of Corollary~\ref{cor:dofgains}}
\label{app:dofgainsproof}
To get $\Delta \mathsf{DoF_{dep}}$, we set the distribution $\mathsf{P}({\cal A})$ in Corollary~\ref{cor:gendofgain} as $p$ if ${\cal A} = \Omega$ and $1-p$ if ${\cal A} = \varnothing$. By some computation, for $KM \leq N$ we get $\Delta \mathsf{DoF_{dep}} = 0$, and for $KM > N$ we get
\begin{align*}
\Delta \mathsf{DoF_{dep}}
& = \min
\left[ \begin{array}{l}
p \min (KM, N+L), p N + (1-p) \min (L, N)
\end{array} \right] - p N \\
& = \min
\left[~ p \min (KM-N, L), (1-p) \min (L, N) ~\right].
\end{align*}

To get $\Delta \mathsf{DoF_{ind}}$, we set the distribution $\mathsf{P}({\cal A})$ in Corollary~\ref{cor:gendofgain} as $\mathsf{B}_{K}(i) = \binom{K}{i} p^i (1-p)^{K-i}$. For notational convenience, we let $i^* := \lfloor \frac{N}{M} \rfloor$. By some computation, we get the claimed gain:
\begin{align*}
\Delta \mathsf{DoF_{ind}} & = \min
\left[ \begin{array}{l}
\sum_{i=0}^{i^*} \mathsf{B}_K(i) \cdot iM + \sum_{i=i^* + 1}^{K} \mathsf{B}_K(i) \min (iM, N+L), \\
\sum_{i=0}^{i^*} \mathsf{B}_K(i) \min (iM+L, N) + \sum_{i=i^*+1}^{K} \mathsf{B}_K(i) \cdot N
\end{array} \right] - \left[ \sum_{i=0}^{i^*} \mathsf{B}_K(i) \cdot iM + \sum_{i=i^*+1}^{K} \mathsf{B}_K(i) \cdot N \right] \\
& = \min
\left[~ \sum\limits_{i=i^*+1}^{K} \mathsf{B}_{K}(i) \min (iM-N, L), \sum\limits_{i=0}^{i^*} \mathsf{B}_{K}(i) \min (L, N-iM) ~\right].
\end{align*}




\section{Proof of Corollary~\ref{cor:necsuf}}
\label{app:necsuf}
We examine if for a certain class of antenna configurations, an upper bound on the sum DoF is strictly less than $K$ times the individual DoF for all $p \in (0, 1)$, as it means the corresponding class is not a necessary condition for attaining collision-free DoF.

If for a certain class of antenna configurations, $K$ times the individual DoF is less than or equal to the sum DoF for $p \in \mathcal{I}$ where $\mathcal{I} \subseteq (0, 1)$, then it means the corresponding class is the necessary and sufficient condition for attaining collision-free DoF, since the individual DoF and the sum DoF are achievable from Theorem~\ref{thm:dofregion}.

\subsection{$KM > N+L$ and $M \leq N$}
\label{appsub:A}
From $M \leq N$, $p\min (M, N+L) = pM$. From $M \leq M+L$ and $M \leq N$, $ pM \leq p \min (M+L, N)$. Thus, we get the individual DoF of $pM$. Using the fact that $\min (a, b) \leq a$, we get an upper bound on the sum DoF: $\sum_{i=0}^{K} {\sf B}_{K}(i) \min (iM, N+L)$.
\begin{align*}
& \sum_{i=0}^{K} {\sf B}_{K}(i) \min (iM, N+L) = \sum_{i=0}^{K-1} {\sf B}_{K}(i) (iM) + {\sf B}_{K}(K) (N+L) \\
& < \sum_{i=0}^{K-1} {\sf B}_{K}(i) (iM) + {\sf B}_{K}(K) (KM) = \sum_{i=0}^{K} {\sf B}_{K}(i) (iM) = K(pM),
\end{align*}
where the last equality is the expectation of a binomial random variable with parameters $K$ and $p$.

In summary, the upper bound on the sum DoF is strictly less than $K$ times the individual DoF for all $p \in (0, 1)$. This class of antenna configurations is \emph{not} a necessary condition for attaining collision-free DoF.

\subsection{$KM > N+L$, $M > N+L$, and $L = 0$}
\label{appsub:B}
From $M > N$ and $L = 0$, we get the individual DoF of $pN$. Using the fact that $\min (a, b) \leq a$ and $L = 0$, we get an upper bound on the sum DoF: $\sum_{i=0}^{K} {\sf B}_{K}(i) \min (iM, N) = \left\{ 1 - (1-p)^K \right\} N$. Let $f(p) := K(pN) - \left\{ 1 - (1-p)^K \right\} N$. Since $f(0) = 0$ and $f'(p) = KN\left\{1 - (1-p)^{K-1} \right\} > 0$ for all $p \in (0, 1)$, $f(p) > 0$ for all $p \in (0, 1)$.

In summary, the upper bound on the sum DoF is strictly less than $K$ times the individual DoF for all $p \in (0, 1)$. This class of antenna configurations is \emph{not} a necessary condition for attaining collision-free DoF.

\subsection{$KM > N+L$, $M > N+L$, and $L \geq 1$}
\label{appsub:C}
From $M > N+L$, $p\min (M, N+L) = p(N+L)$. From $M > N$, $p\min (M+L, N) = pN$. Thus, we get the individual DoF of $
d \leq \min \left\{ p(N+L), pN + (1-p)\min (L, N) \right\}$. Let $p_i := \frac{\min (L, N)}{L + \min (L, N)}$.

For $0 < p < p_i$, the individual DoF is $p(N+L)$. Using the fact that $\min (a, b) \leq a$, we get an upper bound on the sum DoF: $\sum_{i=0}^{K} {\sf B}_{K}(i) \min (iM, N+L) = \left\{ 1 - (1-p)^K \right\} (N+L)$. Using the same method in the earlier case, we can verify that the upper bound on the sum DoF is strictly less than $K$ times the individual DoF for $0 < p < p_i$.

For $p_i \leq p < 1$, the individual DoF is $pN + (1-p)\min (L, N)$. Using the fact that $\min (a, b) \leq b$, we get an upper bound on the sum DoF: $\sum_{i=0}^{K} {\sf B}_{K}(i) \min (iM + L, N) = (1-p)^K \min (L, N) + \left\{ 1 - (1-p)^K \right\} N$. For $p_i \leq p < 1$, $(1-p)^K \min (L, N)$ is strictly less than $K \left\{ (1-p) \min (L, N) \right\}$ and so is $\left\{ 1 - (1-p)^K \right\} N$ than $K(pN)$. In other words, the upper bound on the sum DoF is strictly less than $K$ times the individual DoF for $p_i \leq p < 1$.

In summary, the upper bounds on the sum DoF are strictly less than $K$ times the individual DoF for all $p \in (0, 1)$. This class of antenna configurations is \emph{not} a necessary condition for attaining collision-free DoF.

\subsection{$KM > N+L$, $N < M \leq N+L$, and $L \geq 1$}
\label{appsub:D}
From $M \leq N+L$, $p\min (M, N+L) = pM$.From $N < M$, $p\min (M+L, N) = pN$. Thus, we get the individual DoF of $\min \left\{ pM, pN + (1-p)\min (L, N) \right\}$. When $pM$ is active, by using the same method in Appendix~\ref{appsub:A}, when $pN + (1-p)\min (L, N)$ is active, by using the same method in Appendix~\ref{appsub:C}, we can verify that an upper bound on the sum DoF is strictly less than $K$ times the individual DoF for all $p \in (0, 1)$. This class of antenna configurations is \emph{not} a necessary condition for attaining collision-free DoF.

\subsection{$KM \leq N$}
\label{appsub:E}
From $M < N$, we get the individual DoF of $pM$. From $KM \leq N$, $\min ( iM, N+L ) = iM$ and $iM \leq \min (iM+L, N)$ for all integers $i \leq K$. Thus, we get the sum DoF of $\sum_{i=1}^{K} {\sf B}_{K}(i) iM = K(pM)$. This class of antenna configurations is the necessary and sufficient condition for attaining collision-free DoF for all $p \in (0, 1)$.

\subsection{$N < KM \leq N+L$ and $L \geq 1$}
\label{appsub:F}
From $M < N+L$, $p\min (M, N+L) = pM$. Thus, we get the following individual DoF:
\begin{align*}
d \leq \min \left\{ pM, p\min (M+L, N) + (1-p)\min (L, N) \right\}.
\end{align*}
When $M \leq N$, $pM$ is active for all $p \in (0,1)$. Let $f(p) := pM - \left\{ p\min (M+L, N) + (1-p)\min (L, N) \right\}$. This function is continuous. When $M > N$, since $f(0) < 0$ and $f(1) > 0$, by the intermediate value theorem, there always exists $p_i \in (0, 1)$ such that $f(p_i) = 0$. Thus, for $0 < p < p_i$, $pM$ is active.

From $KM \leq N+L$, $\min ( iM, N+L ) = iM$ for all integers $i \leq K$. Thus, we get the following sum DoF.
\begin{align*}
\sum_{k=1}^{K} d_k \leq \min \left[ K(pM), \sum_{i=0}^{K} {\sf B}_{K}(i) \min (iM+L, N) \right].
\end{align*}
Let $f(p) := K(pM) - \sum_{i=0}^{K} {\sf B}_{K}(i) \min (iM+L, N)$. This function is continuous. Since $f(0) < 0$ and $f(1) > 0$, by the intermediate value theorem, there always exists $p_s \in (0, 1)$ such that $f(p_s) = 0$. Thus, for $0 < p < p_s$, $K(pM)$ is active. One can readily verify that $p_s = \frac{N}{KM}$ when $L \geq N$, and $p_s < \frac{N}{KM}$ when $1 \leq L < N$.

Suppose $p_i < p_s$. Then, for $p_i < p < p_s$, $K$ times the individual DoF of $K \left\{ p\min (M+L, N) + (1-p)\min (L, N) \right\}$ is strictly less than the sum DoF of $K(pM)$. This is a contradiction, since both the individual DoF and the sum DoF are achievable. Thus, $p_s \leq p_i$.

When $M \leq N$, for $0 < p < p_s$, $K$ times the individual DoF of $pM$ is less than or equal to the sum DoF of $K(pM)$. For $p_s \leq p < 1$, the sum DoF of $\sum_{i=0}^{K} {\sf B}_{K}(i) \min (iM+L, N)$ is strictly less than $K$ times the individual DoF of $pM$.

When $M > N$, for $0 < p < p_s$, $K$ times the individual DoF of $pM$ is less than or equal to the sum DoF of $K(pM)$. For $p_s \leq p < p_i$, the sum DoF of $\sum_{i=0}^{K} {\sf B}_{K}(i) \min (iM+L, N)$ is strictly less than $K$ times the individual DoF of $pM$. For $p_i \leq p < 1$, the sum DoF of $\sum_{i=0}^{K} {\sf B}_{K}(i) \min (iM+L, N) = (1-p)^K \min (L, N) + \left\{ 1 - (1-p)^K \right\}N$ is strictly less than $K$ times the individual DoF of $pN + (1-p)\min (L, N)$.

In summary, this class of antenna configurations is the necessary and sufficient condition for attaining collision-free DoF for $p \in (0, p_s)$, where $p_s = \frac{N}{KM}$ when $L \geq N$ and $p_s < \frac{N}{KM}$ when $1 \leq L < N$. 

In conclusion, $KM \leq N+L$ is the necessary and sufficient condition for attaining collision-free DoF for $p \in (0, p^*)$, where $p^* = 1$ if $KM \leq N$, $p^* = \frac{N}{KM}$ if $N < KM \leq N+L$ and $L \geq N$, and $p^* < \frac{N}{KM}$ if $N < KM \leq N+L$ and $1 \leq L < N$.

\section{Proof of~(\ref{depgreaterthanind}): $\Delta \mathsf{DoF_{dep}} > \Delta \mathsf{DoF_{ind}}$}
\label{sec:gainproof}

In this appendix, we perform some mathematical analysis that provides more insight into the discussions in Section~\ref{sec:relaygains} on relaying gains for two representative scenarios: fully dependent and independent user traffic. From the analysis, we prove that $\Delta \mathsf{DoF_{dep}} - \Delta \mathsf{DoF_{ind}} > 0$ when $L \geq \max (KM-N, N)$.

\begin{figure}[!t]
\centering
\includegraphics[width=0.75\columnwidth]{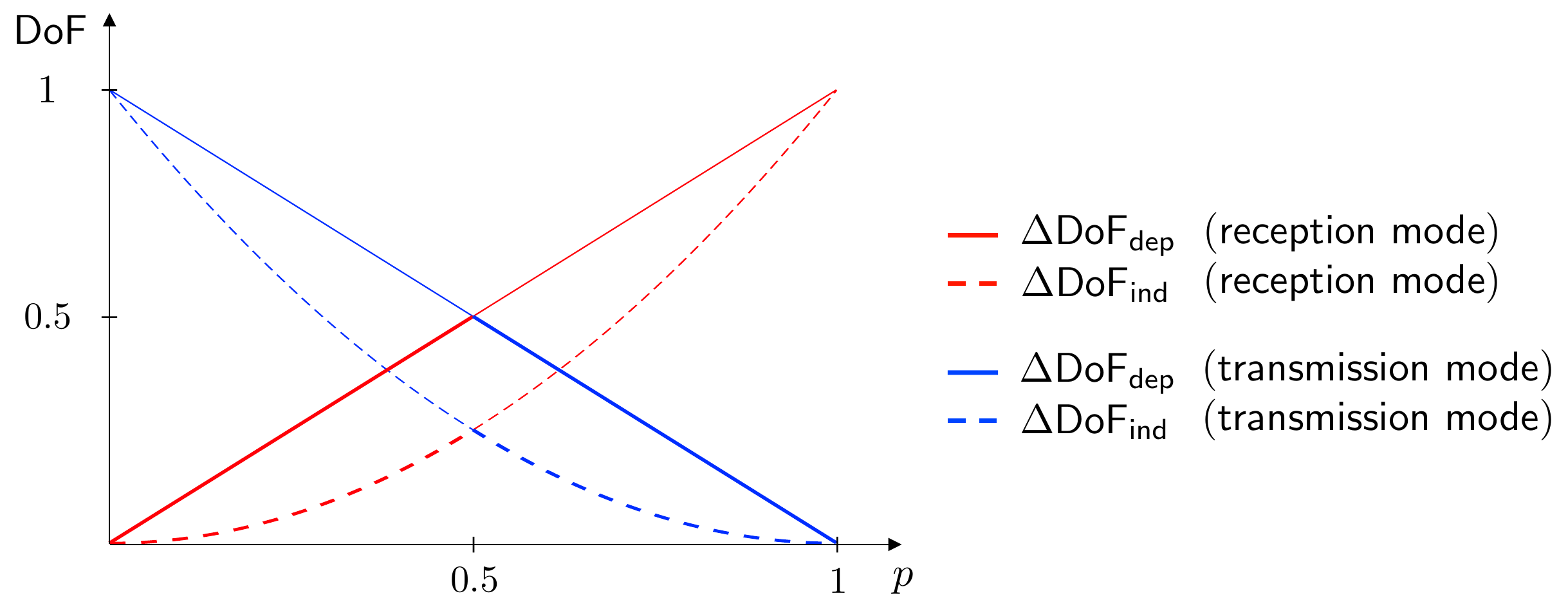}
\caption{The red curves plot the DoF gain obtainable due to the relay's reception capability, and the blue curves plot the DoF gain obtainable due to the relay's transmission capability. For a fixed $p$, the resulting DoF gain attained from the help of a relay (thick curves) is determined by the minimum of the two. For each mode, the DoF gain obtainable for the independent user traffic case (dashed curves) is always below that for the fully dependent user traffic case (solid curves) due to its convexity. It is clear that $\Delta \mathsf{DoF_{dep}} - \Delta \mathsf{DoF_{ind}} > 0$ for $p \in (0, 1)$.}
\label{fig:analysis}
\end{figure}

As explained in Section~\ref{sec:relaygains}, the terms $p \min (KM-N, L)$ in (\ref{eq:depgain}) and $\sum_{i=\lfloor \frac{N}{M} \rfloor+1}^{K} \mathsf{B}_{K}(i) \min (iM-N, L)$ in (\ref{eq:indgain}) quantify the DoF gains that stem from the relay's reception capability for the fully dependent case and independent case, respectively. Also, the terms $(1-p) \min ( L, N )$ in (\ref{eq:depgain}) and $\sum_{i=0}^{\lfloor \frac{N}{M} \rfloor} \mathsf{B}_{K}(i) \min (L, N-iM)$ in (\ref{eq:indgain}) quantify the DoF gains that stem from the relay's transmission capability for the respective two cases. We will show that $\sum_{i=\lfloor \frac{N}{M} \rfloor+1}^{K} \mathsf{B}_{K}(i) \min (iM-N, L)$ is a convex function with respect to $p \in (0, 1)$ that intersects $p \min (KM-N, L)$ at $p = 0, 1$ when $L \geq KM-N$. Similarly, we will show that $\sum_{i=0}^{\lfloor \frac{N}{M} \rfloor} \mathsf{B}_{K}(i) \min (L, N-iM)$ is a convex function with respect to $p \in (0, 1)$ that intersects $(1-p) \min ( L, N )$ at $p = 0, 1$ when $L \geq N$. From the analysis, we can plot Fig.~\ref{fig:analysis}, and prove that $\Delta \mathsf{DoF_{dep}} - \Delta \mathsf{DoF_{ind}} > 0$ when $L \geq \max (KM-N, N)$.

Before we proceed, we obtain the derivative of $\mathsf{B}_{K}(i)$ with respect to $p$. It is helpful for notational convenience.
\begin{align*}
\frac{d}{dp} \mathsf{B}_{K}(i) & \overset{(a)}{=} \frac{d}{dp} \binom{K}{i} p^i (1-p)^{K-i} \\
& \overset{(b)}{=} \binom{K}{i}  i  p^{i-1}(1-p)^{K-i} - \binom{K}{K-i} (K-i) p^i (1-p)^{K-i-1} \\
& = \frac{K}{i} \binom{K-1}{i-1} ip^{i-1}(1-p)^{K-i} - \frac{K}{K-i} \binom{K-1}{K-i-1}(K-i) p^i (1-p)^{K-i-1} \\
& = K \binom{K-1}{i-1} p^{i-1}(1-p)^{K-i} - K \binom{K-1}{K-i-1} p^i (1-p)^{K-i-1} \\
& = K \left\{ \mathsf{B}_{K-1}(i-1) - \mathsf{B}_{K-1}(i) \right\},
\end{align*}
where $(a)$ follows by the definition of $\mathsf{B}_{K}(i)$ in Corollary~\ref{cor:dofgains}, $(b)$ follows by the chain rule of derivatives and the fact that $\binom{K}{i} = \binom{K}{K-i}$, and the rest follows by direct calculations. For completeness, we define $\mathsf{B}_{K-1}(-1) = 0$ and $\mathsf{B}_{K-1}(K) = 0$. It is also helpful to obtain the following for notational convenience.
\begin{align*}
\frac{d}{dp}\sum\limits_{i=x}^{y}\mathsf{B}_{K}(i) & = \left[ K \left\{ \mathsf{B}_{K-1}(x-1) - \mathsf{B}_{K-1}(x) \right\} \right] + \left[ K \left\{ \mathsf{B}_{K-1}(x) - \mathsf{B}_{K-1}(x+1) \right\} \right] \\
& \quad + \cdots + \left[ K \left\{ \mathsf{B}_{K-1}(y-1) - \mathsf{B}_{K-1}(y) \right\} \right] =  K \left\{ \mathsf{B}_{K-1}(x-1) - \mathsf{B}_{K-1}(y)\right\}.
\end{align*}

First, let us analyze the gains from the relay's reception mode. When $L \geq KM-N$, the gains reduce to $p (KM-N)$ and $\sum_{i=\lfloor \frac{N}{M} \rfloor+1}^{K} \mathsf{B}_{K}(i) (iM-N)$. We let $i^* := \lfloor \frac{N}{M} \rfloor$ and $f(p) := \sum_{i=i^*+1}^{K} \mathsf{B}_{K}(i) (iM-N)$. We can modify $f(p)$ as follows:
\begin{align*}
f(p) & = \sum_{i=i^*+1}^{K} \mathsf{B}_{K}(i) (iM-N) \\
& = \sum_{i=i^*+1}^{K} iM \cdot \mathsf{B}_{K}(i) - N \sum_{i=i^*+1}^{K} \mathsf{B}_{K}(i) \\
& = \sum_{i=i^*+1}^{K} iM \cdot \frac{K}{i} p \cdot \binom{K-1}{i-1} p^{i-1} (1-p)^{K-i} - N \sum_{i=i^*+1}^{K} \mathsf{B}_{K}(i) \\
& = KMp \sum_{i = i^* + 1}^{K} \mathsf{B}_{K-1}(i-1) - N \sum_{i=i^*+1}^{K} \mathsf{B}_{K}(i) \\
& = KMp \sum_{i=i^*}^{K-1} \mathsf{B}_{K-1}(i) - N \sum_{i=i^*+1}^{K} \mathsf{B}_{K}(i).
\end{align*}

To see if $f(p)$ is a convex function for $p \in (0, 1)$, we take the second derivative of it. Taking the first derivative, we get:
\begin{align*}
f'(p) & \overset{(a)}{=} KM \sum_{i=i^*}^{K-1} \mathsf{B}_{K-1}(i) + KMp \cdot (K-1) \left\{ \mathsf{B}_{K-2}(i^* -1) - \mathsf{B}_{K-2}(K-1) \right\} - N \cdot K \left\{ \mathsf{B}_{K-1}(i^*) - \mathsf{B}_{K-1}(K) \right\} \\
& \overset{(b)}{=} K(K-1)Mp \cdot \mathsf{B}_{K-2}(i^* - 1) - KN \cdot \mathsf{B}_{K-1}(i^*) + KM \sum_{i=i^*}^{K-1} \mathsf{B}_{K-1}(i),
\end{align*}
where $(a)$ follows by the chain rule of derivatives; and $(b)$ follows by the fact that $\mathsf{B}_{K-2}(K-1) = \mathsf{B}_{K-1}(K) = 0$.

Taking the second derivative, we get:
\begin{align*}
f''(p) & \overset{(a)}{=} K(K-1)M \cdot \mathsf{B}_{K-2}(i^* -1) + K(K-1)Mp \cdot (K-2) \left\{ \mathsf{B}_{K-3}(i^* - 2) - \mathsf{B}_{K-3}(i^* - 1) \right\} \\
& \quad - KN \cdot (K-1) \left\{ \mathsf{B}_{K-2}(i^* - 1) - \mathsf{B}_{K-2} (i^*) \right\} + KM \cdot (K-1) \left\{ \mathsf{B}_{K-2}(i^* - 1) - \mathsf{B}_{K-2}(K-1) \right\} \\
& \overset{(b)}{=} K(K-1)M \cdot \mathsf{B}_{K-2}(i^* -1) + K(K-1)M \left\{ (i^* - 1)  \mathsf{B}_{K-2}(i^* -1) - i^* \mathsf{B}_{K-2}(i^*) \right\} \\
& \quad - K(K-1)N \left\{ \mathsf{B}_{K-2}(i^* - 1) - \mathsf{B}_{K-2} (i^*) \right\} + K(K-1)M \cdot \mathsf{B}_{K-2}(i^* - 1) \\
& = K(K-1) \left\{ (N-i^* M) \mathsf{B}_{K-2}(i^*) - (N-(i^*+1)M) \mathsf{B}_{K-2}(i^*-1) \right\},
\end{align*}
where $(a)$ follows by the chain rule of derivatives; $(b)$ follows by the fact that $p(K-2)\mathsf{B}_{K-3}(i^*-2) = (i^*-1)\mathsf{B}_{K-2}(i^*-1)$ and $p(K-2)\mathsf{B}_{K-3}(i^*-1) = i^*\mathsf{B}_{K-2}(i^*-1)$, and $\mathsf{B}_{K-2}(K-1) = 0$. 

We can see that $f(p)$ is convex for $p \in (0, 1)$ from the fact that $N - i^*M \geq 0$ and $N - (i^*+1)M \leq 0$ due to the definition of $i^* := \lfloor \frac{N}{M} \rfloor$. Also, we can easily verify that at $p = 0, 1$, $f(p)$ intersects $p(KM-N)$.

Now, let us analyze the gains from the relay's transmission mode. When $L \geq N$, the gains reduce to $(1-p) N$ and $\sum_{i=0}^{\lfloor \frac{N}{M} \rfloor} \mathsf{B}_{K}(i) (N-iM)$. We let $i^* := \lfloor \frac{N}{M} \rfloor$ and $f(p) := \sum_{i=0}^{i^*} \mathsf{B}_{K}(i) (N-iM)$. Following a similar series of steps shown earlier, we can modify $f(p)$ as follows:
\begin{align*}
f(p) = N \sum_{i=0}^{i^*} \mathsf{B}_{K}(i) - KMp \sum_{i=0}^{i^* - 1} \mathsf{B}_{K-1}(i).
\end{align*}

Similarly, to see if $f(p)$ is a convex function for $p \in (0, 1)$, we take the second derivative of it. Taking the first derivative, we get:
\begin{align*}
f'(p) & \overset{(a)}{=} N \cdot K \left\{ \mathsf{B}_{K-1}(- 1) - \mathsf{B}_{K-1}(i^*) \right\} - KM \sum_{i=0}^{i^* - 1} \mathsf{B}_{K-1}(i) - KMp \cdot (K-1) \left\{ \mathsf{B}_{K-2}(-1) - \mathsf{B}_{K-2}(i^* - 1) \right\} \\
& \overset{(b)}{=} K(K-1)Mp \cdot \mathsf{B}_{K-2}(i^* - 1) - KN \cdot \mathsf{B}_{K-1}(i^*) - KM \sum_{i=0}^{i^* - 1} \mathsf{B}_{K-1}(i),
\end{align*}
where $(a)$ follows by the chain rule of derivatives; $(b)$ follows by the fact that $\mathsf{B}_{K-1}(-1) = \mathsf{B}_{K-2}(-1) = 0$.

Taking the second derivative, we get:
\begin{align*}
f''(p) & \overset{(a)}{=} K(K-1)M \cdot \mathsf{B}_{K-2}(i^* -1) + K(K-1)Mp \cdot (K-2) \left\{ \mathsf{B}_{K-3}(i^* - 2) - \mathsf{B}_{K-3}(i^* - 1) \right\} \\
& \quad - KN \cdot (K-1) \left\{ \mathsf{B}_{K-2}(i^* - 1) - \mathsf{B}_{K-2} (i^*) \right\} - KM \cdot (K-1) \left\{ \mathsf{B}_{K-2}(- 1) - \mathsf{B}_{K-2}(i^*-1) \right\} \\
& \overset{(b)}{=} K(K-1)M \cdot \mathsf{B}_{K-2}(i^* -1) + K(K-1)M \left\{ (i^* - 1)  \mathsf{B}_{K-2}(i^* -1) - i^* \mathsf{B}_{K-2}(i^*) \right\} \\
& \quad - K(K-1)N \left\{ \mathsf{B}_{K-2}(i^* - 1) - \mathsf{B}_{K-2} (i^*) \right\} + K(K-1)M \cdot \mathsf{B}_{K-2}(i^* - 1) \\
& = K(K-1) \left\{ (N-i^* M) \mathsf{B}_{K-2}(i^*) - (N-(i^*+1)M) \mathsf{B}_{K-2}(i^*-1) \right\},
\end{align*}
where $(a)$ follows by the chain rule of derivatives; $(b)$ follows by the fact that $p(K-2)\mathsf{B}_{K-3}(i^*-2) = (i^*-1)\mathsf{B}_{K-2}(i^*-1)$ and $p(K-2)\mathsf{B}_{K-3}(i^*-1) = i^*\mathsf{B}_{K-2}(i^*-1)$, and $\mathsf{B}_{K-2}(-1) = 0$. 

We can see that $f(p)$ is convex for $p \in (0, 1)$ from the fact that $N - i^*M \geq 0$ and $N - (i^*+1)M \leq 0$ due to the definition of $i^* := \lfloor \frac{N}{M} \rfloor$. Also, we can easily verify that at $p = 0, 1$, $f(p)$ intersects $(1-p)N$.

\section{Proof of increasing $\Delta \mathsf{DoF_{dep}^{peak}}$ and $\Delta \mathsf{DoF_{ind}^{peak}}$}
\label{sec:peakproof}

When $L \geq KM-N$ and $L \geq N$, from Corollary~\ref{cor:dofgains} and some manipulation, we get
\begin{align*}
\Delta \mathsf{DoF_{dep}}
= \min \big( p (KM-N), (1-p) N \big), ~\Delta \mathsf{DoF_{ind}}
= \min ( pKM, N ) - \sum_{i=0}^{K} \mathsf{B}_{K}(i) \min (iM, N).
\end{align*}
It is straightforward to verify that $\Delta \mathsf{DoF_{dep}}$ is maximized at $p^* = \frac{N}{KM}$. To verify that $\Delta \mathsf{DoF_{ind}}$ is also maximized at $p^*$, we consider two cases: $0 < p < \frac{N}{KM}$ and $\frac{N}{KM} \leq p < 1$. For $0 < p < \frac{N}{KM}$, we get
\begin{align*}
\Delta \mathsf{DoF_{ind}} = \sum_{i=\lfloor \frac{N}{M} \rfloor + 1}^{K} \mathsf{B}_{K}(i) (iM-N),
\end{align*}
where the equality holds since $\sum_{i=0}^{K} \mathsf{B}_{K}(i) \cdot iM = pKM$. By taking the derivative of the equality, we can verify that $\Delta \mathsf{DoF_{ind}}$ is increasing.
\begin{align*}
\Delta \mathsf{DoF_{ind}^{'}} = \sum_{i=\lfloor \frac{N}{M} \rfloor + 1}^{K} \frac{\mathsf{B}_{K}(i)}{p(1-p)} (i-pK)(iM-N) \geq 0, 
\end{align*}
where the inequality holds since if $0 < p < \frac{N}{KM}$ and $i \geq \lfloor \frac{N}{M} \rfloor + 1$, then $i - pK \geq 0$ and $iM-N \geq 0$. For $\frac{N}{KM} \leq p < 1$, similarly we can verify that $\Delta \mathsf{DoF_{ind}}$ is decreasing. Both $\Delta \mathsf{DoF_{dep}}$ and $\Delta \mathsf{DoF_{ind}}$ increase for $0 < p < p^*$ and decrease for $p^* \leq p < 1$, thus maximized at $p^* = \frac{N}{KM}$.


Assuming $M=N=1$, from Corollary~\ref{cor:dofgains}, we get
\begin{align*}
\Delta \mathsf{DoF_{dep}^{peak}} = 1 - \frac{1}{K}, ~ \Delta \mathsf{DoF_{ind}^{peak}} = \Big( 1 - \frac{1}{K} \Big)^K.
\end{align*}
We can readily verify that $\Delta \mathsf{DoF_{dep}^{peak}}$ grows as $K$ increases. We can also verify that $\Delta \mathsf{DoF_{ind}^{peak}}$ grows as $K$ increases by showing that the ratio of $\Delta \mathsf{DoF_{ind}^{peak}}$ with $K+1$ users to that with $K$ users is greater than one:
\begin{align*}
\frac{( 1-\frac{1}{K+1} )^{K+1}}{( 1-\frac{1}{K} )^K} = \left( 1-\frac{1}{K} \right) \Big( 1 + \frac{1}{(K+1)(K-1)} \Big)^{K+1} > \left( 1-\frac{1}{K} \right) \left( 1 + \frac{1}{K-1} \right) = 1,
\end{align*}
where the inequality holds due to Bernoulli's inequality. Thus, both $\Delta \mathsf{DoF_{dep}^{peak}}$ and $\Delta \mathsf{DoF_{ind}^{peak}}$ grow as $K$ increases.



Similarly, by applying Bernoulli's inequality to the forward difference of sequence $\Delta \mathsf{DoF_{dep}^{peak}} - \Delta \mathsf{DoF_{ind}^{peak}}$, we can verify that $\Delta \mathsf{DoF_{dep}^{peak}} - \Delta \mathsf{DoF_{ind}^{peak}}$ grows as $K$ increases.





\ifCLASSOPTIONcaptionsoff
  \newpage
\fi



%

\bibliographystyle{IEEEtran}
\bibliography{bib_relayburstymac}

\begin{thebibliography}{10}
\providecommand{\url}[1]{#1}
\csname url@samestyle\endcsname
\providecommand{\newblock}{\relax}
\providecommand{\bibinfo}[2]{#2}
\providecommand{\BIBentrySTDinterwordspacing}{\spaceskip=0pt\relax}
\providecommand{\BIBentryALTinterwordstretchfactor}{4}
\providecommand{\BIBentryALTinterwordspacing}{\spaceskip=\fontdimen2\font plus
\BIBentryALTinterwordstretchfactor\fontdimen3\font minus
  \fontdimen4\font\relax}
\providecommand{\BIBforeignlanguage}[2]{{%
\expandafter\ifx\csname l@#1\endcsname\relax
\typeout{** WARNING: IEEEtran.bst: No hyphenation pattern has been}%
\typeout{** loaded for the language `#1'. Using the pattern for}%
\typeout{** the default language instead.}%
\else
\language=\csname l@#1\endcsname
\fi
#2}}
\providecommand{\BIBdecl}{\relax}
\BIBdecl

\bibitem{jafar:it09}
V.~R. Cadambe and S.~A. Jafar, ``Degrees of freedom of wireless networks with
  relays, feedback, cooperation, and full duplex operation,'' \emph{IEEE
  Transactions on Information Theory}, vol.~55, no.~5, pp. 2334--2344, May
  2009.

\bibitem{kim:isit15}
S.~Kim, I.-H. Wang, and C.~Suh, ``A relay can increase degrees of freedom in
  bursty interference networks,'' \emph{IEEE International Symposium on
  Information Theory}, June 2015.

\bibitem{Meulen:it71}
E.~C. van~der Meulen, ``Three-terminal communication channels,'' \emph{Advances
  in Applied Probability}, vol.~3, pp. 120--154, 1971.

\bibitem{Cover:it79}
T.~M. Cover and A.~A. El-Gamal, ``Capacity theorems for the relay channel,''
  \emph{IEEE Transactions on Information Theory}, vol.~25, pp. 572--584, Sept.
  1979.

\bibitem{kramer:it05}
G.~Kramer, M.~Gastpar, and P.~Gupta, ``Cooperative strategies and capacity
  theorems for relay networks,'' \emph{IEEE Transactions on Information
  Theory}, vol.~51, no.~9, pp. 3037--3063, Sept. 2005.

\bibitem{noisy:it11}
S.~H. Lim, Y.-H. Kim, A.~{El Gamal}, and S.-Y. Chung, ``Noisy network coding,''
  \emph{IEEE Transactions on Information Theory}, vol.~57, no.~5, pp.
  3132--3152, May 2011.

\bibitem{salman:it11}
S.~Avestimehr, S.~Diggavi, and D.~Tse, ``Wireless network information flow: A
  deterministic approach,'' \emph{IEEE Transactions on Information Theory},
  vol.~57, no.~4, pp. 1872--1905, Apr. 2011.

\bibitem{Kramer:allerton04}
G.~Kramer, ``Models and theory for relay channels with receive constraints,''
  \emph{42nd Annual Allerton Conference on Communication, Control, and
  Computing}, Sept./Oct. 2004.

\bibitem{ElGamal:it07}
A.~{El Gamal}, N.~Hassanpour, and J.~Mammen, ``Relay networks with delays,''
  \emph{IEEE Transactions on Information Theory}, vol.~53, no.~10, Oct. 2007.

\bibitem{Abramson:ALOHA}
N.~Abramson, ``The {ALOHA} system --- {A}nother alternative for computer
  communications,'' \emph{Proceedings of AFIPS Conference}, pp. 281--285, Nov.
  1970.

\bibitem{kleinrock:comm75}
L.~Kleinrock and F.~A. Tobagi, ``Packet switching in radio channels: Part {I}
  --- {C}arrier sense multiple-access modes and their throughput-delay
  characteristics,'' \emph{IEEE Transactions on Communications}, vol.~23,
  no.~12, pp. 1400--1416, Dec. 1975.

\bibitem{khude:isit09}
N.~Khude, V.~Prabhakaran, and P.~Viswanath, ``Opportunistic interference
  management,'' \emph{IEEE International Symposium on Information Theory}, pp.
  2076--2080, June/July 2009.

\bibitem{wang:spawc13}
I.-H. Wang and S.~Diggavi, ``Interference channels with bursty traffic and
  delayed feedback,'' \emph{IEEE International Workshop on Signal Processing
  Advances in Wireless Communications}, June 2013.

\bibitem{wang:isit13}
I.-H. Wang, C.~Suh, S.~Diggavi, and P.~Viswanath, ``Bursty interference channel
  with feedback,'' \emph{IEEE International Symposium on Information Theory},
  July 2013.

\bibitem{wang:isit14}
S.~Mishra, I.-H. Wang, and S.~Diggavi, ``Harnessing bursty interference in
  multicarrier systems with feedback,'' \emph{IEEE International Symposium on
  Information Theory}, July 2014.

\bibitem{gamalkim:nit}
A.~{El Gamal} and Y.-H. Kim, \emph{Network Information Theory}.\hskip 1em plus
  0.5em minus 0.4em\relax Cambridge University Press, 2011.

\bibitem{mohajer:it11}
S.~Mohajer, S.~Diggavi, and D.~Tse, ``Approximate capacity of a class of
  {G}aussian interference-relay networks,'' \emph{IEEE Transactions on
  Information Theory}, vol.~57, no.~5, pp. 2837--2864, May 2011.

\end{thebibliography}

%
%

%

%
%
%




\end{document}